\documentclass{jpp_}

\usepackage[utf8]{inputenc}
\usepackage{latexsym}
\usepackage{graphics}
\usepackage{graphicx}
\usepackage{subcaption}
\usepackage{amsmath}
\usepackage{amssymb}
\usepackage{color}
\usepackage[T1]{fontenc}
\usepackage[english]{babel}
\newcommand{\ie}{\emph{i.e., }}
\newcommand{\exg}{\emph{e.g., }}
\newcommand{\omp}{\omega_p}
\newcommand{\reff}[1]{(\ref{#1})}
\newcommand{\eref}[1]{Eq.\reff{#1}}
\newcommand{\erefs}[1]{Eqs.\reff{#1}}
\newcommand{\figref}[1]{Fig.\ref{#1}}

\newcommand{\emphpanel}[1]{{#1}}

\newcommand{\deltat}{\Delta \tau}

\shorttitle{Resonance overlap and non-linear features of the BPS}
\shortauthor{N. Carlevaro, G. Montani, M.V. Falessi}

\title{Resonance overlap and non-linear features \\ of the beam-plasma system}

\author{N. Carlevaro\aff{1,2}
\corresp{\email{nakia.carlevaro@enea.it}},
G. Montani\aff{1,3} \and
M.V. Falessi\aff{1,4}}

\affiliation{
\aff{1}ENEA, Fusion and Nuclear Safety Department, C. R. Frascati,\\ Via E. Fermi 45, 00044 Frascati (Roma), Italy
\aff{2}Consorzio RFX, Corso Stati Uniti 4, 35127 Padova, Italy
\aff{3}Physics Department, ``Sapienza'' University of Rome, P.le Aldo Moro 5, 00185 Roma, Italy
\aff{4}INFN - Rome section, P.le Aldo Moro 2, 00185 Roma, Italy
}

\begin{document}
\maketitle

\begin{abstract}
The beam-plasma instability can be addressed as a reduced model in several contexts of plasma physics, from space to fusion plasma. In this paper, we review and refine some non-linear features of this model. Specifically, by analyzing the dependence of the non-linear velocity spread as a function of the linear growth rate, we discuss the effective size of the resonance in view of its role in the spectral overlap at saturation. The relevance of this characterization relies on the necessity of a quantitative determination of the overlap degree to discriminate among different transport regimes of the self consistent dynamics. The analysis is enriched with a study of the phase-space dynamics by means of the Lagrangian Coherent Structure technique, in order to define the transport barriers of the system describing the relevant features of the overlap process. Finally, we discuss relevant features related to the mode saturation levels.
\end{abstract}


\section{Introduction}
The beam-plasma interaction is one of the most interesting paradigms of plasma physics \citep{ZCrmp}, not only for its direct implementation in laboratory physics, \exg for plasma accelerators \citep{Li14,ES96,KJ86}, but also because it is isomorphic to the bump-on-tail problem \citep{BS11,shalaby17,pommo17}. In fact, the relaxation of supra-thermal particle beams provides a paradigm for the quasi-linear theory of weak plasma turbulence \citep{Vedenov,Pines}, finding several applications from fusion plasma to astrophysics and cosmic geophysics. In particular, the non-linear interaction between resonant particles and electrostatic waves (and the corresponding hole and clump mechanism) is addressed to analyze the generation of whistler-mode chorus and electromagnetic ion cyclotron emissions in space plasmas (see \exg \citet{tzc17} and \citet{tobita18} and references therein). Moreover, the phenomenology of particle trapping in the Langmuir potential well has been proposed as a reduced model for the behavior of fast ions interacting with Alfv\'en waves in fusion devices \citep{BB90a,BB90b,BB90c,BS11}. 

The main features of the beam-plasma system (BPS) are the dispersion relation (characterizing the linear phase of instability) and the non-linear evolution of the mode, \exg the saturation of fluctuating fields and particle trapping \citep{OM68,OWM71,MK78,TMM94,BB95a}. The BPS can be cast as a $N$-body scheme for both the thermal distribution of plasma electrons (assuming a neutralizing ion background) and the supra-thermal particles constituting the beam itself \citep{EEbook,ee18}. However, in \citet{OWM71}, the problem was successfully reduced assuming that the thermal plasma could be described as linear dielectric medium, in which the beam electrons interact with a single Langmuir mode. This approach has been generalized to the case of a warm electron beam (in which the initial velocity dispersion is significant) in \citet{L72}, and the implications of this paradigm for the transport features (convection and diffusion) of beam electrons have been analyzed in  \citet{BB95b,VK12,ncentropy}. Recent applications of this theoretical framework to interpret fast-ion transport in Tokamaks can be found in \citet{nceps16,nceps19}.

The basic features of the interaction of a tenuous beam with a cold plasma have been extensively discussed. In this paper, we review and refine by means of numerical simulations some standard findings, adding original details regarding the phase-space trapping dynamics related to the non-linear spread of particles in overlapping and non-overlapping regimes.

We start by discussing the relevant scalings with respect to the system drive, in the case of a single mode (see the pioneering works of \citet{ucla_fried,OWM71,L72,BB90a,wu94}). The analysis is developed for a realistic initial profile of the beam distribution function. We outline the quadratic behavior of the saturated mode amplitude with respect to the linear growth rate and the linear scaling of both the trapping frequency and the estimate for the non-linear velocity spread of the beam particles. We also measure the forming clump behavior in the phase space (due to the saturated mode evolution), in order to characterize the resonance width. This analysis is based on the clustering of particle trajectories in the trapping region of the potential well. We show that the mean size of this region linearly scales with the drive of the system.

We then address the behavior of a three-resonance scenario. Instead of discussing the onset of the overlap as a function of the drive (\citet{L72,BB94a}), we analyze the features of the non-linear velocity spread as responsible for the mode overlap. We thus fix the density parameter of the beam in order to obtain the minimal condition for the overlapping regime at saturation. We show how the measured clump size would not be predictive in defining the overlap, but an additional factor of about 1.3 is required. When the distribution function is considered, this redefined non-linear velocity spread well corresponds to the effective distortion of the profile, and it differs from the clump size, which seems to determine only the ``plateau'' region. This reflects the fact that also particles that are not trapped by the wave (as discussed, for example, in \citet{EEbook}) are relevant in the ``active'' overlap of different non-linear fluctuations, since the power transfer also involves those particles simultaneously feeling two (or multiple) electric fields. Through this mechanism, the resonant wave-particle power exchange can be enhanced. 

With our analysis, we characterize the non-linear size of the resonance in a fully self-consistent scheme, where the distribution function is dynamically coupled to the spectral evolution. Such self-consistency allows one to take into account the backreaction induced by the distribution function deformation on the fluctuation spectrum which is a crucial point in identifying the real resonance region involved in the non-linear dynamics. In this respect, this analysis enters the longstanding debate about the transition to stochasticity of adjacent resonances associated with the well-known Chirikov overlap criterion \citep{Ch60,Ch79} and its deepening and upgradings \citep{Gr68,JL72,ED81,BB95b,LL10}. Actually, having a quantitative characterization of the degree of overlap in the presence of multiple resonances is crucial in determining different transport scenarios of the BPS (in particular, discriminating between pure diffusive and convective transport in the velocity space). This tool takes an important role when the BPS is implemented as a reduced scenario.

A detailed description of the phase-space dynamics is also provided by means of the Lagrangian Coherent Structure (LCS) technique (see \citet{Haller_2015} for a complete review). This approach is particularly suited as a post-processing technique for describing particle motion in the phase space in the presence of a time-dependent scalar potential and it has been already applied to the BPS in \citet{CFMZJPP}. The value of calculating LCSs in this work stems from the fact that they generalize dynamical structures observed in autonomous and periodic systems, \exg invariant manifolds, to temporally aperiodic flows. We will consider only hyperbolic LCSs that organize the flow by attracting or repelling phase-space volume elements over a finite time span and, for the sake of simplicity, we will refer to these specific lines in the phase space as LCS. As already shown in \citet{CFMZJPP}, by means of this technique we can describe the shape of the clumps enclosing trapped particles and thus characterize resonance overlap. Consequently, the phase space at each instant can be partitioned into different subdomains with small or negligible exchange of particles between them. In the case of single resonance, the LCS plots can be compared with the size of the region in the velocity space involved in the particle power exchange as discussed above.

We finally discuss the relevant feature of the BPS by which overlapping resonances yield enhanced fluctuation levels at saturation, with respect to the isolated resonance case. This behavior is a consequence of an efficient transfer of phase-space energy to the Langmuir modes \citep{BB95a}. It takes place only in the presence of adjacent resonances that are weakly overlapped, and it is a relevant physical process also for fast-ion transport induced by Alfv\'en modes in Tokamaks (see, for instance, the recent analysis in \citet{spb16}). With the help of a toy model, we also show the mechanism responsible for this phenomenon which relies on the enhancement of available free energy in overlapping regimes, as outlined in \citet{BB95a}.

The paper is structured as follows. In Sec.\ref{bps}, the adopted BPS equations are described and the basic simulation parameters are given. The particle trapping mechanism is discussed and the description of the non-linear velocity spread is introduced also in terms of the clump size. Trapping frequency, mode saturation level and non-linear velocity spread are studied by means of numerical simulations as a function of the linear drive. In Sec.\ref{overlap}, the issue of resonance overlap is addressed, and the effective the non-linear velocity spread is defined using a scale factor applied to the clump width. In Sec.\ref{ftle}, the LCS technique is applied to describe phase-space dynamics during the overlap. In Sec.\ref{modes}, the mode saturation level is analyzed. Concluding remarks are given in Sec.\ref{conclusion}.

\section{Hamiltonian description of the beam-plasma interaction}\label{bps}
The BPS describes the non-linear dynamics of a fast electron beam interacting with the a background plasma in a 1-dimensional approximation. The dynamics is regulated by the Vlasov-Poisson equation, when collisions can be neglected (\ie the collision frequency is much smaller than the plasma frequency). Such a system describes the coupled dynamics of the electric field and the electron distribution function and its analytical treatment is limited by the non-linear character of the resulting dynamics. For a description in terms of a graph theory, see \citet{AK66}. However, when addressing the numerical study of the Vlasov-Poisson system, a significant simplification is offered by the Hamiltonian approach \citep{OWM71,EEbook}. In this representation, the electron distribution function is reduced to a discrete set of beams (actually, on a numerical level, such an issue is conceptually mandatory) and then the self-consistent evolution of the Newton's second law for the electrons and of the Fourier-transformed Poisson equation for the electric field (generated by the charge displacement) is naturally addressed in terms of dimensionless universal quantities. From a numerical point of view, this approach is less demanding with respect to the direct integration of the Vlasov equation. As a result, a large number of resonances can be easily described, up to dealing with the so-called quasi-linear limit \citep{BB95b,VK12}. The conceptual framework that justifies the implementation of a Hamiltonian approach is the so-called notion of ``quasi-stationary states'' \citep{BB95a,EEbook,CFGGMP14}, \ie transient states of the discrete system, in which a time is spent by the dynamics proportional to the number of considered particles (in practice, charged macroparticles). Such states approach for a diverging number of particles, \ie in the continuum limit, the equilibrium configuration of the Vlasov-Poisson system. 

In this work, following \citet{OM68} and \citet{OWM71} (see also \citet{CFMZJPP,ncentropy} for other reference details), the plasma is addressed as a cold linear dielectric medium represented by longitudinal electrostatic Langmuir waves, whose density $n_p$ is assumed much greater that the beam density $n_B$. In this respect, we define $\eta\equiv n_B/n_p$ as one of the fundamental parameters of the model. The Langmuir wave is described in terms of the corresponding modes whose frequencies $\omega$ are very close to the plasma frequency $\omp$: the dielectric function (for a fixed mode) $\epsilon=1-\omp^2/\omega^2$ is, thus, nearly vanishing. This allows one to write an evolutive equation for the modes corresponding to the Poisson equation. The simple force equation governs, instead, the particle dynamics. A single mode is set to be resonantly excited by the beam, considering wave number $k=\omega/v_{r}$, where $v_{r}$ is a fixed initial resonant velocity of the beam particles.

In this scheme, the cold background plasma is considered as a periodic slab of length $L$, and particle positions are labeled by $x_i$ ($i=1,\,...,\,N$, where $N$ indicates the total particle number). The Langmuir wave scalar potential $\varphi(x,t)$ is addressed by means of the Fourier components $\varphi_k(t)$ and we use the following dimensionless quantities: $\bar{x}_i=x_i(2\pi/L)$, $\tau=t\omp$, $u_i=\bar{x}_i'=v_i(2\pi/L)/\omp$, $\ell=k(2\pi/L)^{-1}$, $\phi_\ell=(2\pi/L)^2 e\varphi_\ell/m\omp^2$, $\bar{\phi}_\ell=\phi_\ell e^{-i\tau}$. The derivative with respect $\tau$ is indicated with the prime, while barred frequencies (and the related growth rates) are normalized as $\bar{\omega}=\omega/\omp$ ($\bar{\gamma}=\gamma/\omp$). The governing equations of the BPS reads 
\begin{align}\label{mainsys1}
\begin{split}
&\bar{x}_i'=u_i \;,\\
& u_i'=\sum_{\ell}\big(i\,\ell\;\bar{\phi}_\ell\;e^{i\ell\bar{x}_{i}}+c.c.\big)\;,\\
&\bar{\phi}_\ell'=-i\bar{\phi}_\ell+\frac{i\eta}{2\ell^2 N}\sum_{\ell} e^{-i\ell\bar{x}_{i}}\;,
\end{split}
\end{align}
while the resonance condition is $\ell=\bar{\omega}/u_{r}$.

As reference case for the analysis of this work, we consider a positive slope as initial warm beam distribution function in the velocity space:
\begin{align}\label{erffb}
F_0(u)=0.5\;\textrm{Erfc}[a-b\,u]\;,
\end{align}
with the beam distributed from $u_{min}=0.001$ to $u_{max}=0.002$ (with $a\simeq6.8$ and $b\simeq4537$). In the non-linear simulations, we initialize $N=10^{6}$ particles and implement a Runge-Kutta (fourth-order) algorithm to solve \erefs{mainsys1}. Actually, the initialization in the velocity space is formal and particles are free to spread in this coordinate, while we consider uniform initial conditions for the particle positions between $0$ and $2\pi$ and we apply periodic boundary conditions in this range for $\bar{x}_i$. Specifically, we discretize the beam profile in 500 delta-like beams, each of them having a single fixed velocity. Each beam is also initialized with random generated positions and with a number of particles provided by \eref{erffb} (normalized to $N$). This procedure provides the initial $2\times N$ array for $(\bar{x}_i,u_i)$ which is evolved in time self-consistently with the modes. The particles are thus considered as points in the phase space and no $(\bar{x},u)$-meshes are required. The modes are instead initialized at $\mathcal{O}(10^{-14})$ (this guarantees the initial linear regime). For the considered time scales, both the total energy and momentum (for the explicit expressions, see \citet{CFMZJPP}) are conserved with relative fluctuations of about $1.4\times10^{-5}$.

We conclude by writing down the the linear dispersion relation for electric field perturbations. Using dimensionless variables, it reads \citep{OM68,LP81}
\begin{align}\label{disrel}
2(\bar{\omega}_0+i\bar{\gamma}_L-1)-\frac{\eta}{\ell M}
\int_{-\infty}^{+\infty}\!\!\!\!\!\!\!du\frac{\p_u F_0(u)}{u\ell-\bar{\omega}_0-i\bar{\gamma}_L}=0\;,
\end{align}
where $M=\int F_0(u)du$ and we have explicitly written $\bar{\omega}=\bar{\omega}_0+i\bar{\gamma}_L$. Here, $\bar{\omega}_0$ includes a small real frequency shift with respect to $\omp$ and $\bar{\gamma}_L$ indicates the normalized linear growth rate \citep{nceps18}. This expression describes the inverse Landau damping mechanism, and defines the linear instability condition of a single mode through the resonance pole $u=\bar{\omega}_0/\ell$.

\subsection{Non-linear velocity spread}
Let us now review some basic non-linear features of the BPS in the single-mode assumption, by means of numerical simulations of \erefs{mainsys1}. The dynamics of one isolated unstable mode consists of an initial exponential growth (characterized by $\bar{\gamma}_L$) followed by non-linear saturation, where particles are trapped and begin to slosh back and forth in the potential well of the wave. This makes the mode intensity oscillate and generates rotating clumps in the phase space. In \citet{BB95a} (and references therein), also verified was the existence of two distinct saturation regimes in the presence of source and sink. They correspond, on the one hand, to the previously discussed steady-state saturation and, on the other hand, to the so-called pulsating scenario in the correspondence of large mode damping rate (not addressed in the present paper).

A quadratic relation exists \citep{OWM71,L72,ucla_fried} between the saturation level of the considered linearly unstable mode (dubbed $|\bar{\phi}|^{S}$) and the linear growth rate, \ie $|\bar{\phi}|^{S}=\alpha\bar{\gamma}_L^{2}$ (with $\alpha=const.$). This relation holds only if the non-linear dynamics is not sensitive to the morphology of the distribution function \citep{ZCrmp,nceps16}; and all the studies reported in the present work satisfy this condition.

Assuming a single-mode scheme, the approximation of the post-saturation dynamics by an instantaneous harmonic oscillator allows one to identify the so-called trapping (bounce) frequency $\omega_B$ as
\begin{align}\label{ombeq}
\bar{\omega}_B=\sqrt{2\ell^{2}|\bar{\phi}|^{S}}=
\sqrt{2\alpha}\;\ell\,\bar{\gamma}_L\;.
\end{align}
Meanwhile, from energy conservation at saturation, one can estimate the non-linear velocity spread of resonant particles, \ie particles having velocity $u_r=1/\ell$ (here and in the following, we approximate $\bar{\omega}_0=1$). This quantity is clearly related to the (half) rotating clump width mentioned above and it is derived from the relation $m(\Delta\tilde{v}_{NL})^{2}/2=e|\varphi(x,t)|^{S}$. Using dimensionless variables, this definition of the non-linear velocity spread can be cast as
\begin{align}\label{estnlvs}
\Delta\tilde{u}_{NL}/u_r=2\ell\sqrt{|\bar{\phi}|^{S}}=
\sqrt{2}\,\bar{\omega}_B=2\ell\sqrt{\alpha}\;\bar{\gamma}_L\;.
\end{align}
This estimate does not account for effects of non-resonant particles. Thus, in order to get a satisfactory characterization of the non-linear dynamics, we introduce the clump width $\Delta{u}^{c}_{NL}$ defined by measuring the maximum instantaneous velocity of particles initialized at $\tau=0$ with $u<u_r$; and, similarly, the minimum velocity of particles initialized with $u>u_r$. This corresponds monitoring the instantaneous spread of particles above and below resonance. Such a measure is performed during the temporal evolution of the system and $\Delta{u}^{c}_{NL}$ is taken as the value at saturation time $\tau_S$.
\begin{figure}
\centering
\includegraphics[width=0.48\textwidth]{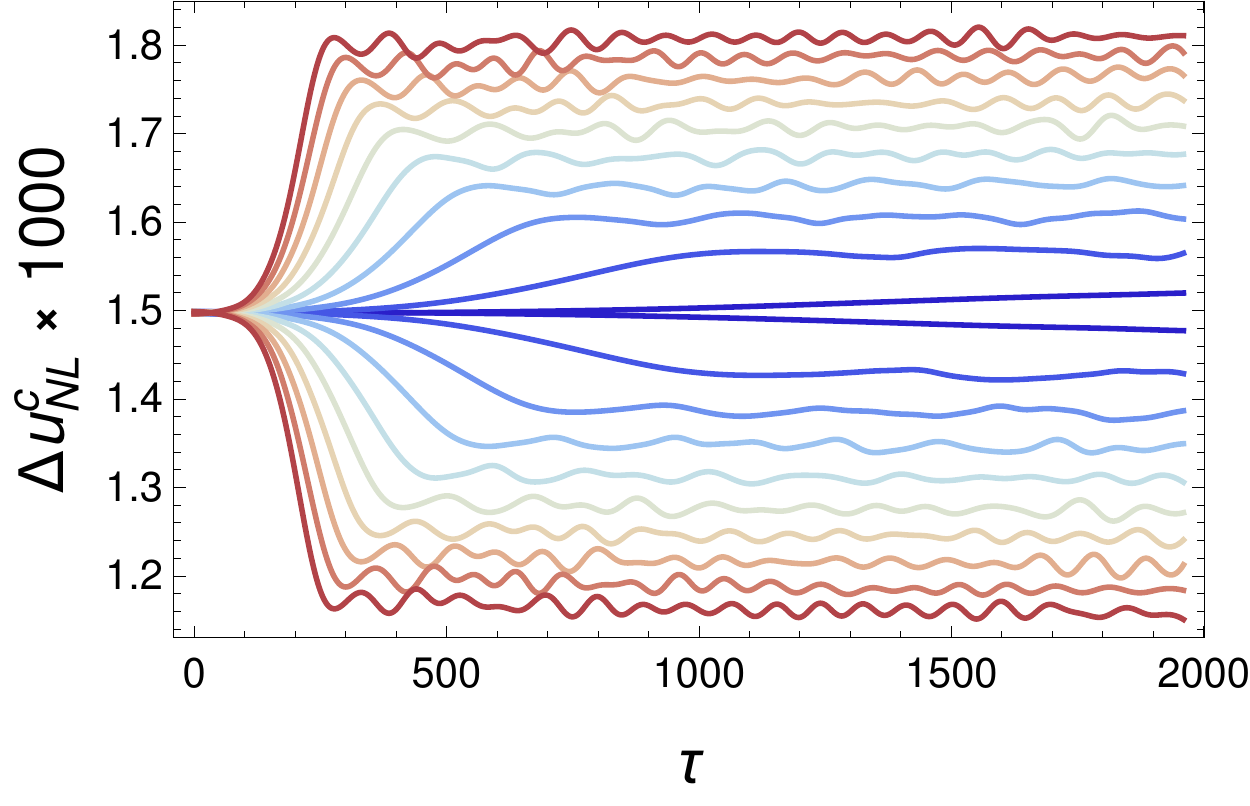}
\includegraphics[width=0.48\textwidth]{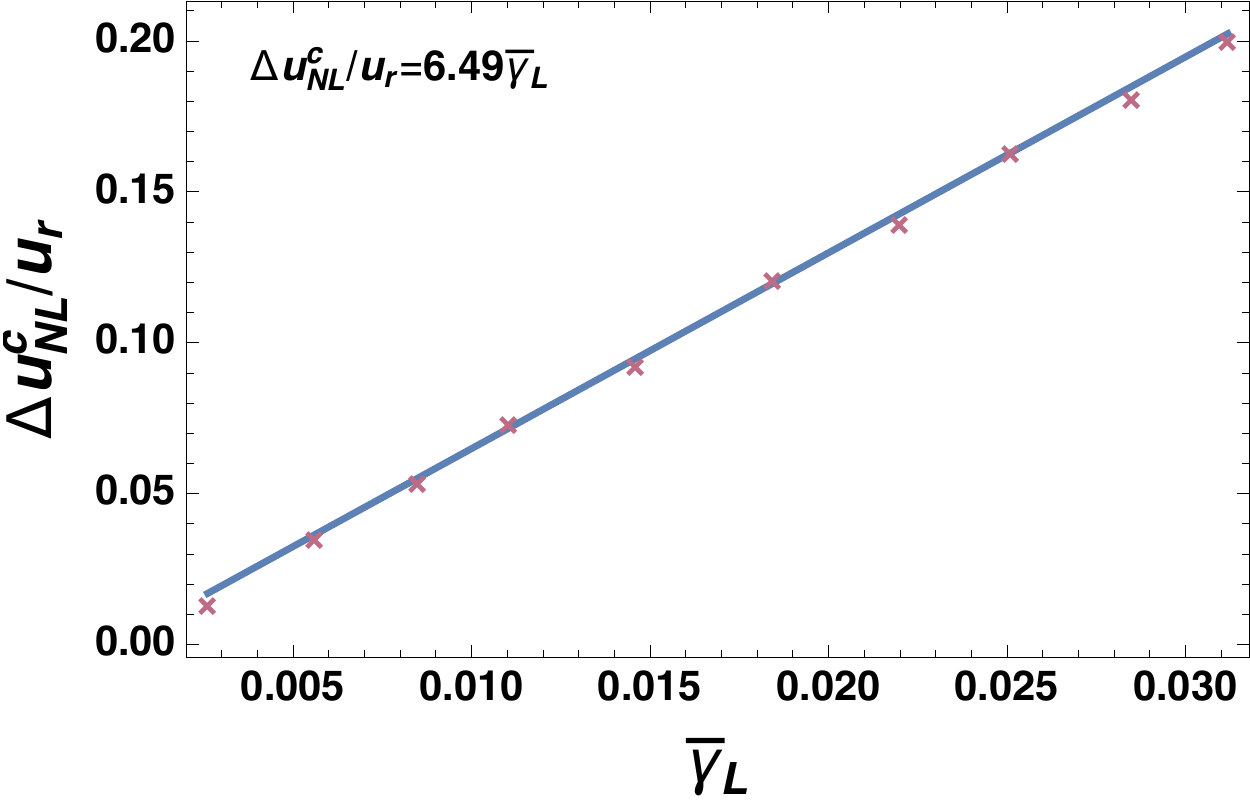}
\caption{(Color online) Case $u_r\simeq0.0015$: crosses represent simulation results while solid lines are linear fits. \emphpanel{Left-hand panel}: Plot of the clump width $\Delta{u}^{c}_{NL}$ as a function of $\tau$. Different colors represent the 10 equispaced values of $\eta\in[0.00015\,\textrm{(blue)},\,0.0025\,\textrm{(red)}]$. \emphpanel{Right-hand panel:} Dependence of $\Delta{u}^{c}_{NL}$ as a function of $\bar{\gamma}_L$.
\label{fig_poug}}
\end{figure}

Let us discuss the behavior of the quantities introduced above, as a function of the linear growth rate. The following analysis refines well-known results reported in the literature, mainly derived for linear shapes of the initial distribution profile \citep{ucla_fried,OWM71,L72,wu95,wu94}, enlightening the crucial role of wave-particle trapping for non-linear mode saturation and resonance broadening. We study 5 distinct cases having different resonant velocities (namely $u_r\simeq0.0013,\,0.0014,\,0.0015,\,0.0016,\,0.0017$). For each case, 10 simulations with different $\eta$ values are studied (equispacing the $\eta$ value from $0.00015$ to $0.0025$), providing distinct drives $\bar{\gamma}_L$ by means of \eref{disrel}, and then linear fits are implemented to estimate the scalings. We obtain the following behaviors
\begin{align}
|\bar{\phi}|^{S}&=1.27\pm0.38\times10^{-5}\;\bar{\gamma}_L^{2}\;,\\
\bar{\omega}_B&=3.31\pm0.07\;\bar{\gamma}_L\;,\\
\Delta\tilde{u}_{NL}/u_r&=4.72\pm0.1\;\bar{\gamma}_L\;,\\
\Delta u^{c}_{NL}/u_r&=6.64\pm0.15\;\bar{\gamma}_L\;.
\end{align}

Regarding the measured clump width, for the sake of completeness, we show the details of the case at the inflection point of $F_0$ ($u_r\simeq0.0015$). In the \emphpanel{left-hand panel} \figref{fig_poug}, we plot the clump width $\Delta{u}^{c}_{NL}$ versus time and for the different values of $\eta$ (different colors in the figure): as expected, the smaller the value of $\eta$, the smaller is  $\Delta{u}^{c}_{NL}$ is. In fact, as the value of $\eta$ is lowered, the instability drive becomes correspondingly weaker, and, in turn, the electric field amplitude at saturation and the clump width are decreased. We observe that, for a fixed $\eta$ value, the clump width increases with time during the linear instability growth phase, until the saturation level is reached. In the \emphpanel{right-hand panel} of \figref{fig_poug}, we instead illustrate the behaviors of $\Delta{u}^{c}_{NL}$ as a function of the linear drive outlining the linear scaling.

We emphasize that repeating the analysis above with different slope of the Erfc distribution, \ie varying, at a given $u_r$ value, the slope of the distribution function $\p_u f_B(u)$ (responsible for change of $\bar{\gamma}_L$), yields quantitatively comparable behaviors. This suggests a universal character of the scalings.

\vspace{5mm}
\section{Resonance overlap at saturation}\label{overlap}
In this Section, we discuss the problem of resonance overlap. Such an issue is often faced by characterizing the onset of the overlap in terms of the drive (\exg \citet{L72,BB94a}). This means that a threshold is determined in the linear growth rate in order to discriminate between overlapping and isolated regimes. Such a threshold can be estimated directly by enhancing the instability drive, or by characterizing the phase velocity distance between neighboring modes. In the following, we discuss the parallel issue of a proper characterization of the non-linear velocity spread as responsible for the mode overlap, also in view of the analysis of the phase-space dynamics of the next Section. In this sense, we set an overlapping system in the correspondence of the threshold drive and we analyze the predictivity of the already discussed non-linear velocity spread evaluations.

\begin{figure}
\centering
\includegraphics[width=0.48\textwidth]{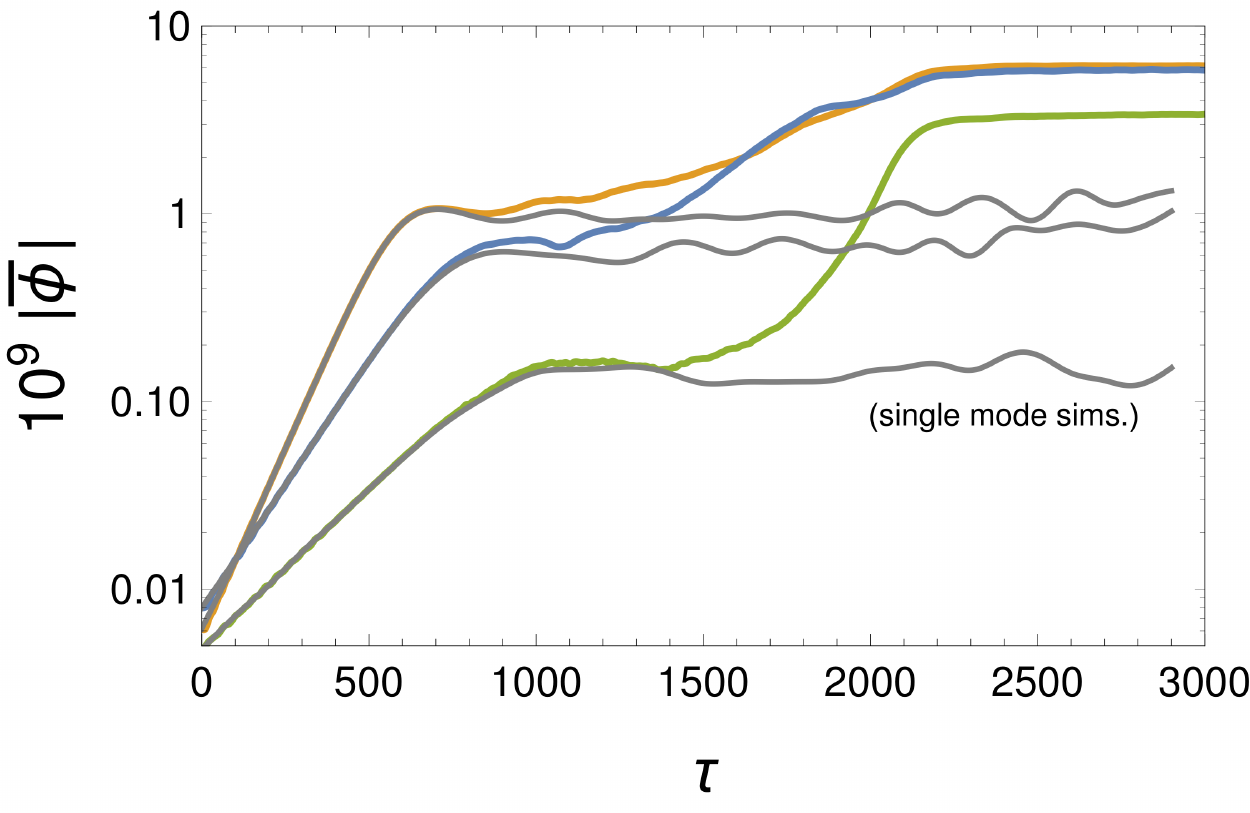}
\includegraphics[width=0.46\textwidth]{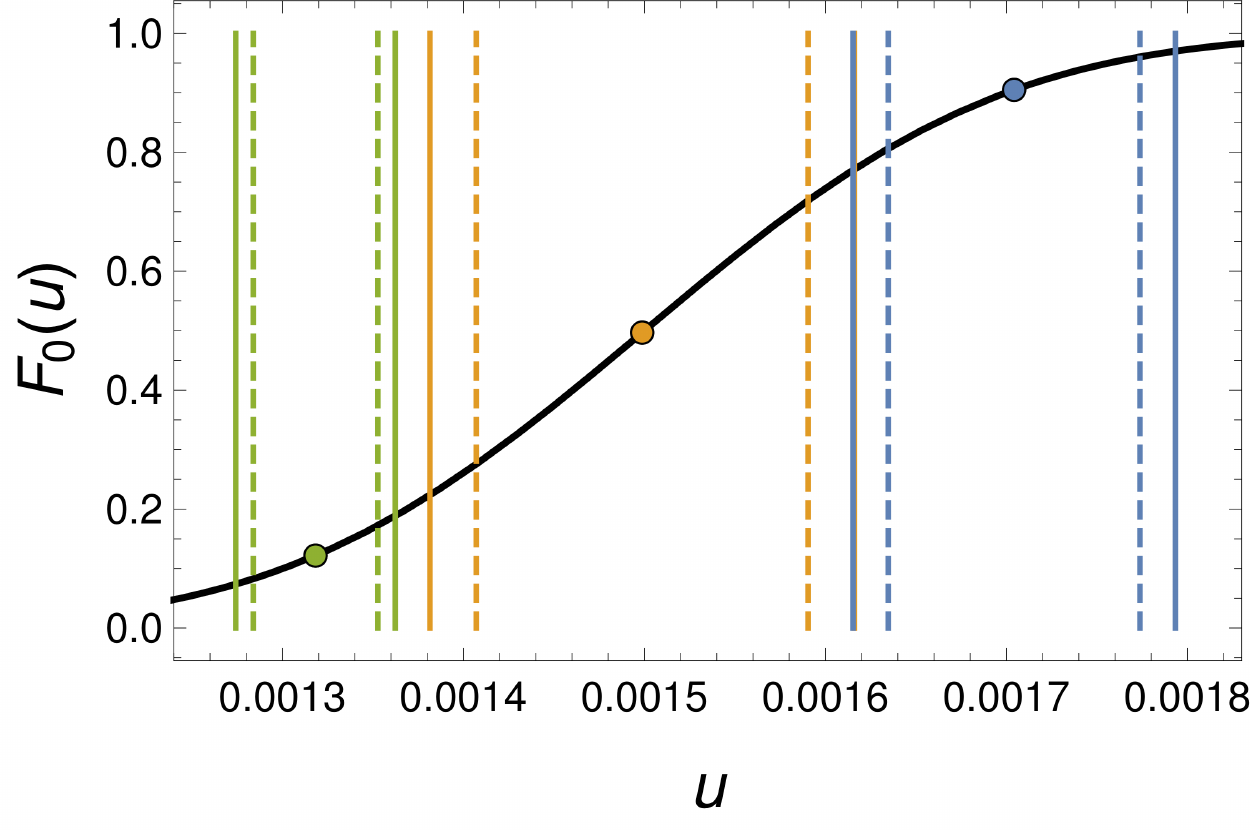}
\caption{(Color online) \emphpanel{Left-hand panel}: Plot of the mode evolution in a $3$ resonance model (colored lines) and the respective single-mode simulations (gray lines), for $\eta=0.00055$. The color scheme is: $u_{r1}\simeq0.0013$ (green), $u_{r2}\simeq0.0015$ (yellow), $u_{r3}\simeq0.0017$ (blue). \emphpanel{Right-hand panel}: Three resonance system. The black line represents $F_0$ of \eref{erffb}. Circles denote $u_{r1}$ (green), $u_{r2}$ (yellow) and $u_{r3}$ (blue). We indicate also $u_r\pm\Delta u^{c}_{NL}$ (dashed lines) and $u_r\pm\Delta u_{NL}$ (solid lines) with corresponding colors.
\label{fig_over}}
\end{figure}

We set a system in which $3$ distinct modes are excited in the correspondence of different resonant velocities ($u_{r1}\simeq0.0013$, $u_{r2}\simeq0.0015$ and $u_{r3}\simeq0.0017$). We consider that resonance overlap occurs when the phase-space regions associated with different resonances mix, due to the non-linear velocity spread. In the considered case, the onset of the overlap regime at saturation time emerges for $\eta\geqslant0.00055$, as clearly represented by the mode evolution depicted in the \emphpanel{left-hand panel} of \figref{fig_over}. The system is evolved self-consistently for the $3$ modes and it is compared with the single-mode simulation results of each resonance (gray lines). As it can be argued from the plot, two resonances start to interact nearby the corresponding single-mode saturation time. In particular, by increasing the value of $\eta$, the resonances become increasingly more overlapped and the interaction time becomes smaller. On the contrary, reducing the drive results in a progressive separation of the resonances, which eventually behave as isolated. For small $\eta$, however, some residual non-linear interplay is found, although the overlap starts much later than single-mode saturation time.

Let us now depict the resonance position and the corresponding $\Delta u^{c}_{NL}$ (\emphpanel{right-hand panel} of \figref{fig_over}, dashed lines) for the threshold value $\eta=0.00055$. In this case, the non-linear trapping regions appear, actually, non-overlapped, suggesting that fluctuations should evolve as superposition of non-interacting modes. Such an evidence has the physical implication that in the ``active'' overlap of different non-linear fluctuations, also particles that are not trapped by the wave \citep{EEbook} are relevant. This phenomenon is related to the analysis of the transition to stochasticity of adjacent resonances. There is vast a literature on this specific topic and the corresponding well-known Chirikov overlap criterion \citep{Ch60,Ch79} (for details, see also \citet{ED81,BB95a,LL10}). It is worth remarking that in the present work, the fluctuating spectrum is not imposed, as usual \citep{EEbook,LL10}, as an external field. It is instead self-consistently determined including also the coupled beam particle non-linear evolution, as in the already mentioned works related to the threshold drive for the onset of overlapping. According to the literature, \exg \citet{LL10}, the quantity $\Delta u^{c}_{NL}$ must be enlarged by means of a scale factor to obtain the observed overlap of the resonance width. In particular, taking into account \figref{fig_over} (right-hand panel), we have to multiply the clump widths evaluated using $\Delta u^{c}_{NL}=6.64 \bar{\gamma}_L$ (dashed lines) by a factor $>1$ in order to obtain a non-zero intersection of the new re-sized non-linear spread regions (solid lines). In our case, this is represented by the yellow/blue line intersection around $u=0.00162$. We can thus recognize an effective non-linear velocity spread $\Delta u_{NL}$, able to properly describe the verified resonance regime, as
\begin{align}\label{new_du}
\Delta u_{NL}=\beta\Delta u^{c}_{NL}\;,\qquad\beta\simeq1.28\;,
\end{align}
well represented in the figure with solid lines.

\begin{figure}
\centering
\includegraphics[width=0.48\textwidth]{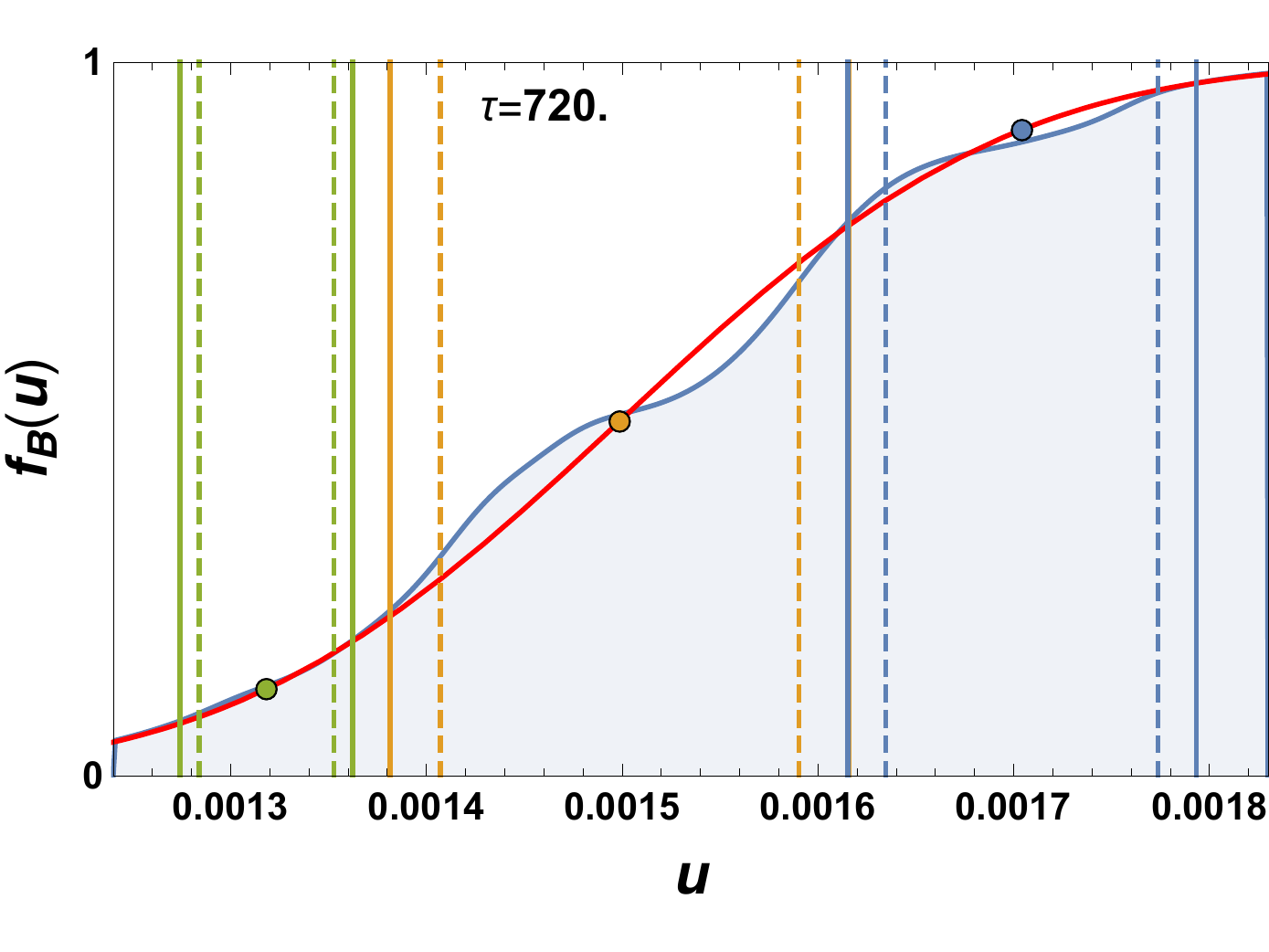}
\includegraphics[width=0.48\textwidth]{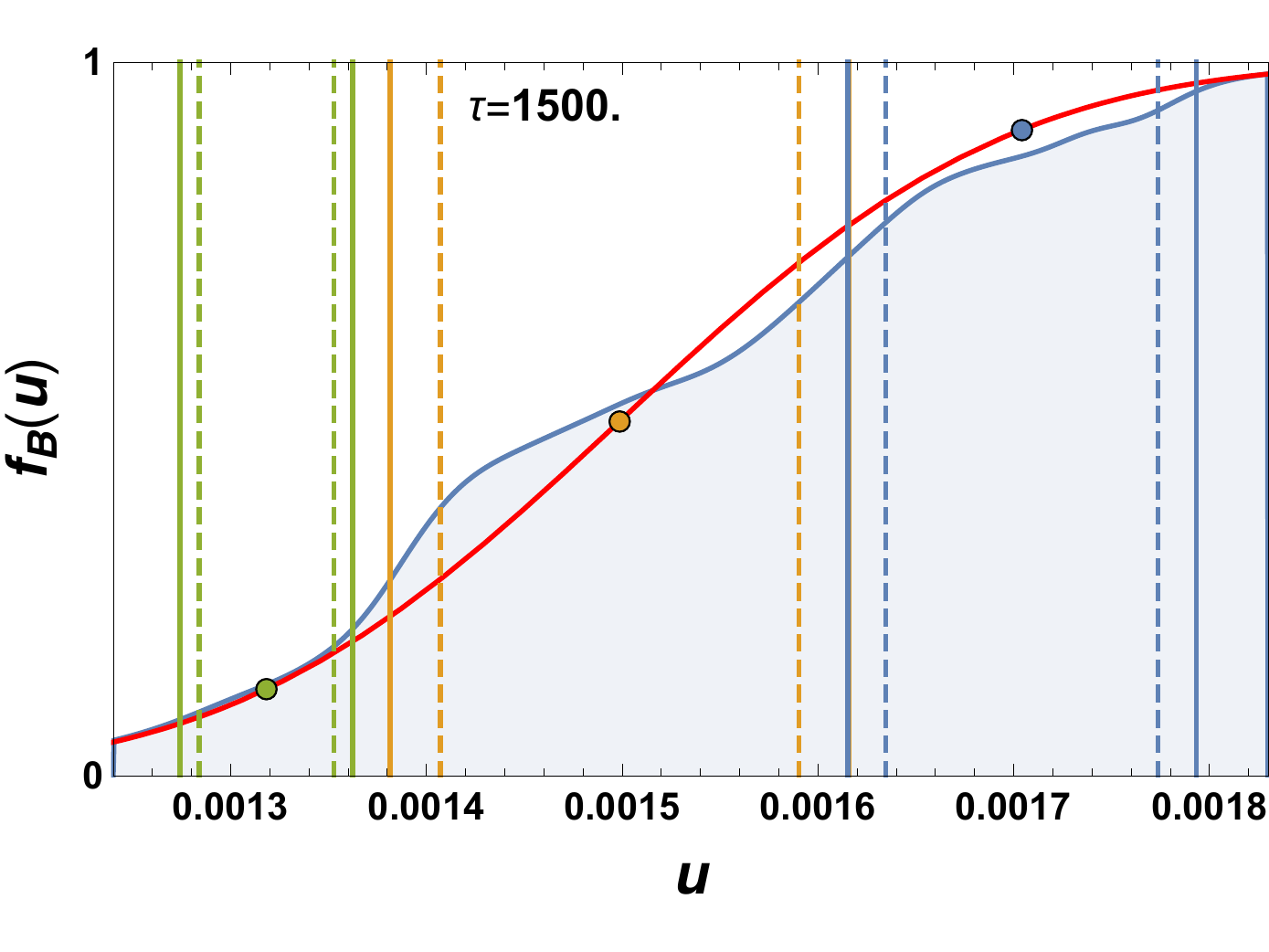}
\caption{(Color online) Beam distribution function at saturation $f_B(u,\tau_S=720)$ (left-hand panel) and at a later instant $f_B(u,\tau=1500)$ (right-hand panel) for $\eta=0.00055$ in the case of 3 resonances. Color scheme and other notations as in \figref{fig_over} (right-hand panel). The red line represents $F_0$.\label{fig_nlnlnl_}}
\end{figure}

It is worth observing that, for the obtained $\beta$ value, the resonance $u_{r1}$ (represented in green) results in being initially isolated even using the redefined non-linear width. Nonetheless, in this scenario, the non-linear interplay of the overlapping resonances ($u_{r2}$ and $u_{r3}$) affects the dynamics by broadening the corresponding region of non-linear velocity spread (as discussed in \citet{BB94a,BB94b,BB95a}). This can be recognized in the evolution of the distribution function $f_B(u,\tau)$ plotted in \figref{fig_nlnlnl_}. The initial merging at saturation time of the second two resonances (yellow and blue) is evident, while for later times the morphology of the distribution profile starts to affect the resonance region of the first (green) resonance. In this way, the $u_{r1}$ overlap sets in at this slightly later time (cf. also the \emphpanel{left-hand panel} of \figref{fig_over}), after the other fluctuations have been non-linearly amplified. This eventually allows synergistic interaction also with the initially isolated resonance. We also mention that, setting a system in which only the first two resonances $u_{r1}$ (green) and $u_{r2}$ (yellow) are considered, it can be shown how (for the sake of simplicity, we do not propose other plots) they properly behave as two independent modes as predicted by the quantity $\Delta u_{NL}$ in \eref{new_du}.

A careful analysis of the BPS is developed in \citet{BB95a}, where different regimes of the dynamics are considered and numerical simulations are shown to be in agreement with the theoretical models \citep{BB92PRL,BB93,BB94a,BB94b}, \exg the pulsating regime. The present re-analysis wants to emphasizes the role played by intrinsic non-linear effects, due to the self-consistent evolution of the system. Actually, the clump size is typically fixed via analytical estimates, which break down the self-consistency by assuming a frozen field. In \citet{finelli19}, it has been clearly shown how retaining this self-consistency is of crucial importance to ensure predictivity. Here, we also argue how the overlapping process of two or more resonances can strictly depend on small deformations of the distribution function in those regions of the velocity space which are out of the plateau. Such apparently marginal deformations contribute to the real size of the resonant region and play a crucial role when the BPS is implemented into the dynamics of fast ions in the Alfv\'en spectrum of a fusion device \citep{BB90a,BB90b,BB90c}.

\begin{figure}
\centering
\includegraphics[width=0.48\textwidth]{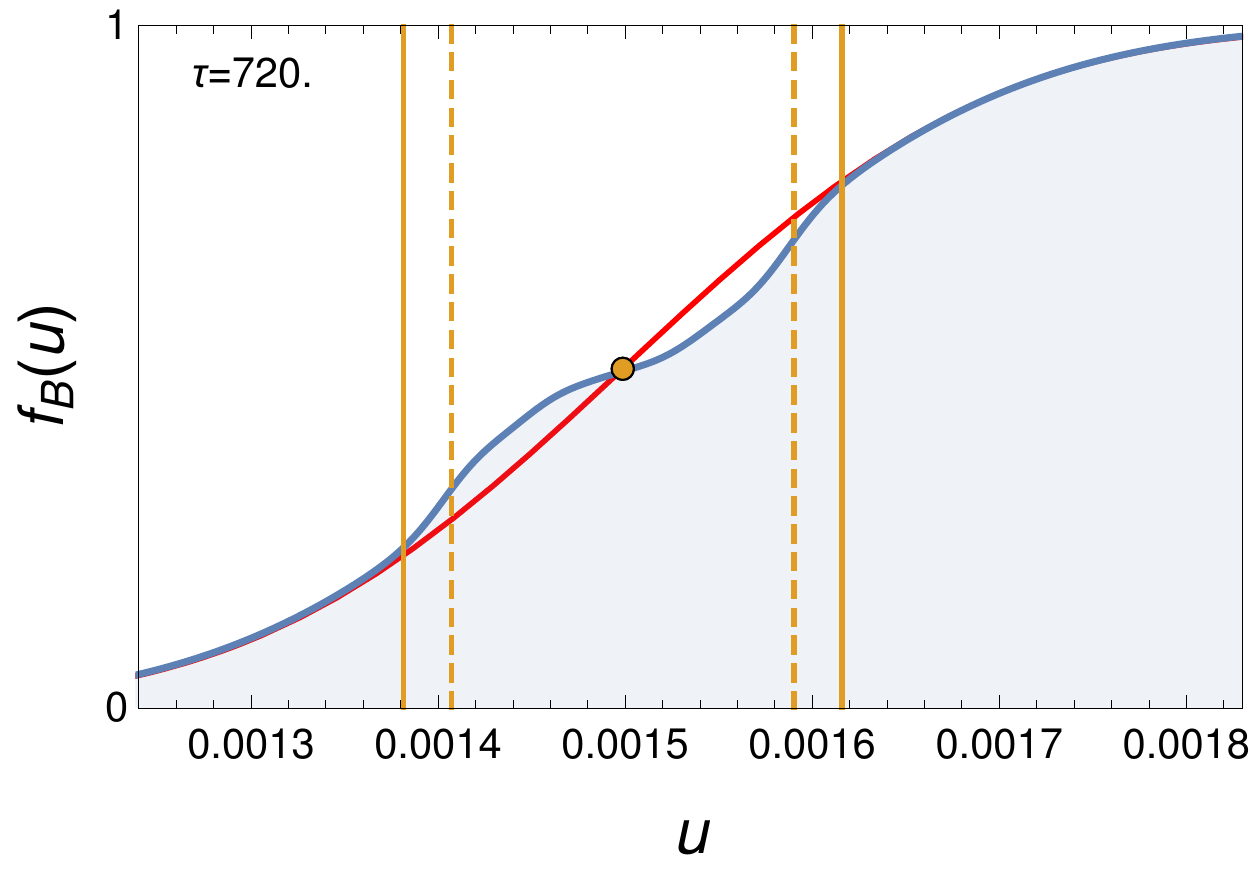}
\caption{(Color online) Snapshot of the beam distribution function at saturation $f_B(u,\tau_S=720)$ for $\eta=0.00055$ in the presence of one single resonance $u_{r2}$ ,represented by the circle (red line represents $F_0$). We also indicate $u_r\pm\Delta u^{c}_{NL}$ (dashed lines) and $u_r\pm\Delta u_{NL}$ (solid lines).\label{fig_nlnlnl}}
\end{figure}

In view of the study of the next Section, we conclude by analyzing the morphology of the beam distribution function in the presence of only the single resonance $u_{r2}$ (the saturation time is set $\tau_S=720$). The results are summarized in \figref{fig_nlnlnl}. The role of un-trapped particles is clear in this scheme: the re-sized non-linear velocity spread $\Delta u_{NL}$ well defines the global distortion of $f_B(\tau_S)$ with respect to the initial profile $F_0$, including un-trapped but nearly resonant particles. These are represented by the plateau edges, which is relevant in the active overlap.

The present study provides an example regarding the difficulty of having quantitative predictions for the degree of overlap based on the linear parameters (the system drive). The physical impact of this fact relies on the possibility of discriminating or excluding specific transport regimes directly from the linear set-up in order to apply (or not) reduced schemes for the dynamics. The focus on the saturation time is due to the necessity of applying the standard estimate for the particle trapping dynamics. Moreover, this choice is also natural because at the saturation time the field amplitude approaches a constant value and we can compare this picture with the case of an assigned external field. Only slightly different results could be obtained by setting later times and/or extending the time scale for the mixing of the phase space to occur. 

We conclude this Section by stressing how this analysis of the beam-plasma instability refines in a fully-self consistent fashion (also using more realistic profiles for the particle distribution) the already existing studies regarding the transition to the stochasticity made by dynamical system theory. In particular, using the scaling addressed in the previous Section, we get 
\begin{align}
\Delta u_{NL}/u_r\simeq8.5\;\bar{\gamma}_L\;.
\end{align}
This expression provides a prescription about the resonance width directly from linear information and it is able to properly describe the resonance overlap regimes, improving the estimates of the trapping region related to the mode saturation amplitude.

\section{Lagrangian Coherent Structure analysis}\label{ftle}
Let us now describe phase-space dynamics produced by resonance overlap using the LCS technique, in comparison to the distribution profile evolution of \figref{fig_nlnlnl_} and \figref{fig_nlnlnl}. LCSs are a generalization of dynamical structures observed in autonomous and periodic systems, \exg invariant manifolds, to temporally aperiodic flows and allow one to identify phase-space regions distinguished by qualitatively different behavior of particle motion (see \citet{Haller_2015} or \citet{MPJPP,CFMZJPP,Di_Giannatale_2018, Di_Giannatale_2018b,Pegoraro_2019} for applications to plasma physics). Following the work of \citet{Haller_2015}, we define LCSs as the most repulsive or attractive material lines (1-dimensional ensemble of material points, \ie points advected by the dynamics) with respect to the nearby ones and, therefore, they are associated with peaked profiles of the Finite Time Lyapunov Exponent (FTLE) fields. 

In order to calculate the FTLE fields, we trace several test particle trajectories under the action of the time-dependent scalar potential generated from an $N$-body simulation. Test particles are initialized to sample the whole phase space of interest, at a fixed time \(\tau\), in two phase-space grids having an infinitesimal displacement in the velocity direction. By other words, a test particle located in \((\bar{x},u)\) has a neighbor initialized in \((\bar{x},u+\delta_\tau)\). Evolving such a system with assigned time-dependent potentials, at a time \(\tau+\deltat\) these two test particles will be at a distance \(\delta_{\deltat}\) in the phase space and the FTLE value \(\sigma\) in the point \((\bar{x},u)\) can be evaluated using the following expression:
\begin{equation}\label{eqftle}
\sigma(\bar{x},u,\tau,\deltat)=
\ln\,(\delta_{\deltat}/\delta_\tau)/\deltat\;.
\end{equation}
When considering a positive time span  \(\deltat>0\), the curves where the FTLE field is peaked define repulsive transport barriers, while when setting \(\deltat<0\) they represent attractive ones. The LCS can be visualized by plotting the maximum values of \(\sigma(\bar{x},u,\tau,\deltat)\) as extracted from a contour plot in the phase space. In the analysis of this Section, we have set two grids of \(200\times200\) test particles thus obtaining \(4000\) values of the FTLE for each phase-space snapshot. We apply the methodology introduced in \citet{CFMZJPP} and we choose a relatively small value for $\Delta \tau$ which highlights finite-time transport barriers instead of the system asymptotic properties such as invariant manifolds. The scalar electrostatic potentials have been sampled from the complete $N$-body simulations.
\begin{figure}
\centering
\includegraphics[width=0.48\textwidth]{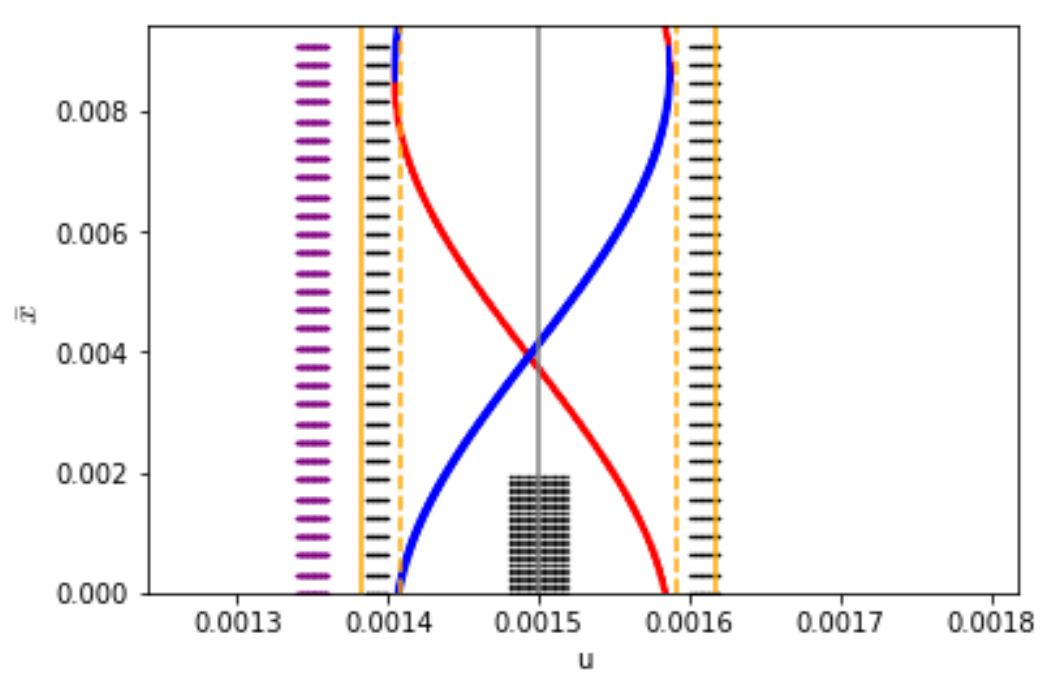}
\includegraphics[width=0.48\textwidth]{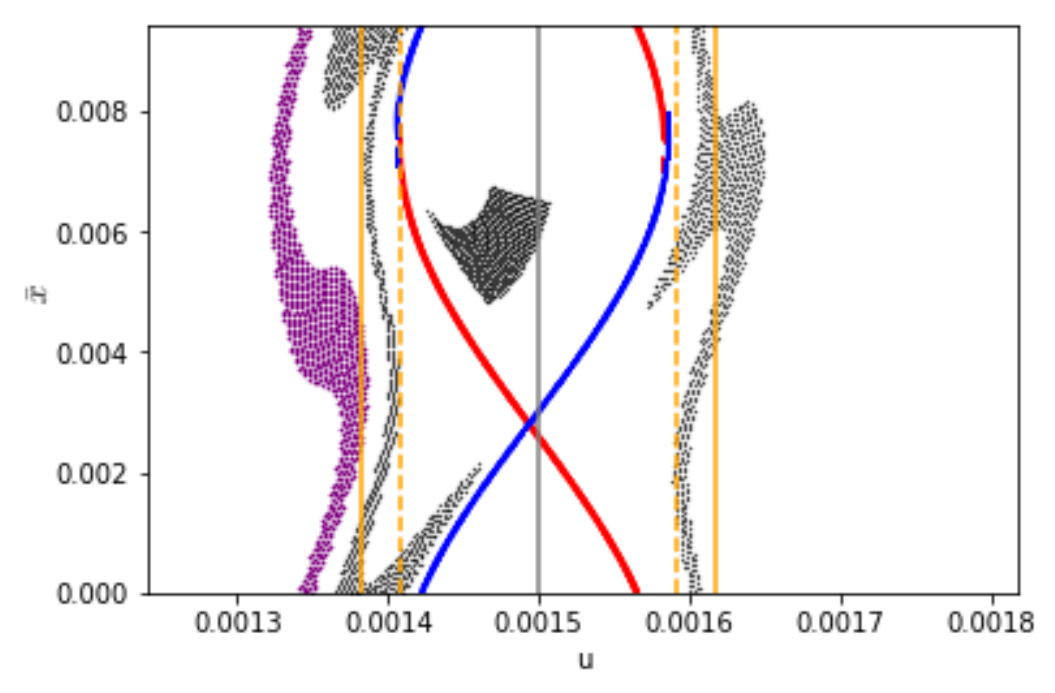}
\caption{(Color online) Attractive and repulsive LCSs (in blue and red, respectively) in the presence of one resonant mode ($u_r\simeq0.0015$), marked by a gray line, calculated by means of the FTLE field with $\Delta \tau = 20$ at $\tau = 700$ (\emphpanel{left-hand panel}) and $\tau=800$ (\emphpanel{right-hand panel}). Dashed lines are placed at $u_{r} \pm \Delta u^{c}_{NL}$ while solid ones at $u_{r} \pm \Delta u_{NL}$ (cf. \figref{fig_nlnlnl}). Passive tracers are depicted with different colors to show phase-space dynamics. 
\label{fig_singlemodeFTLE}}
\end{figure}
\begin{figure}
\centering
\includegraphics[width=0.48\textwidth]{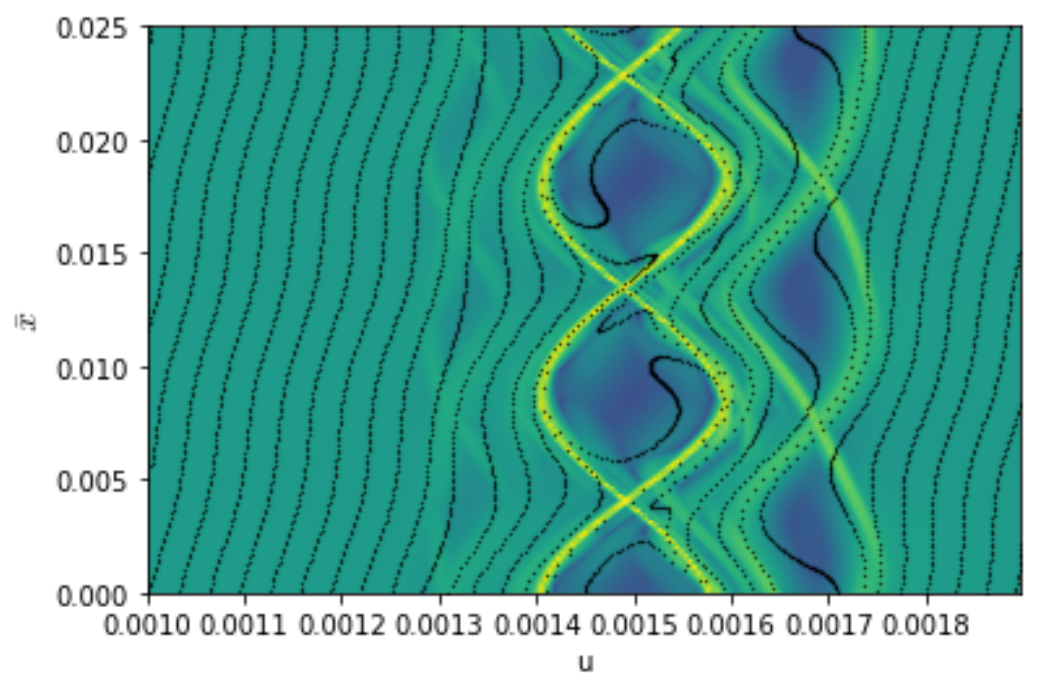}
\includegraphics[width=0.48\textwidth]{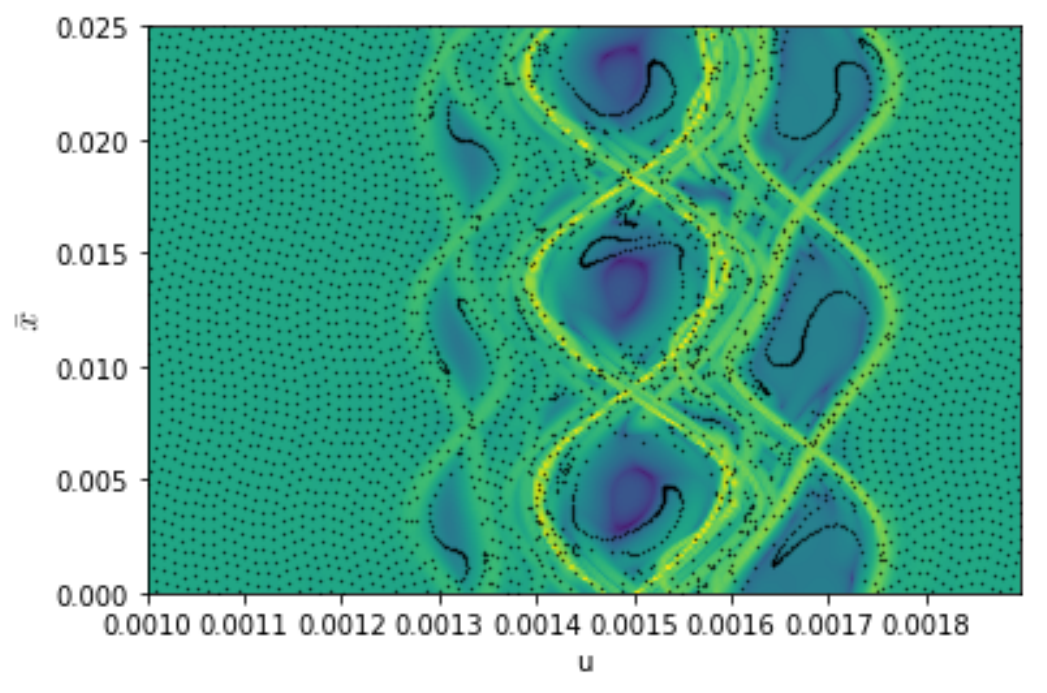}\\
\includegraphics[width=0.48\textwidth]{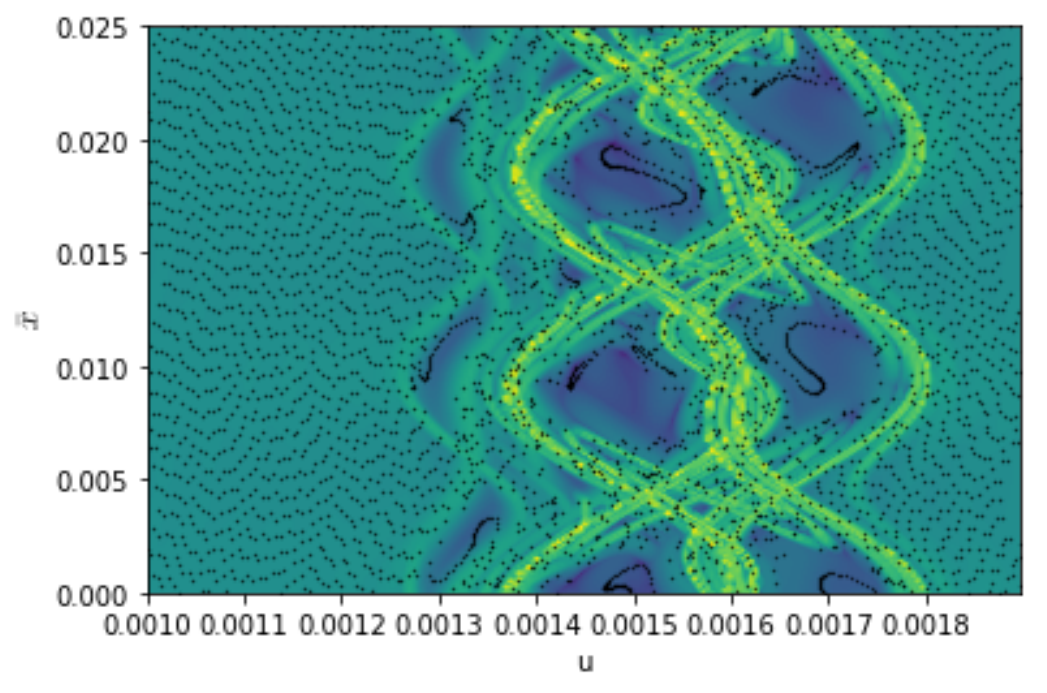}
\includegraphics[width=0.48\textwidth]{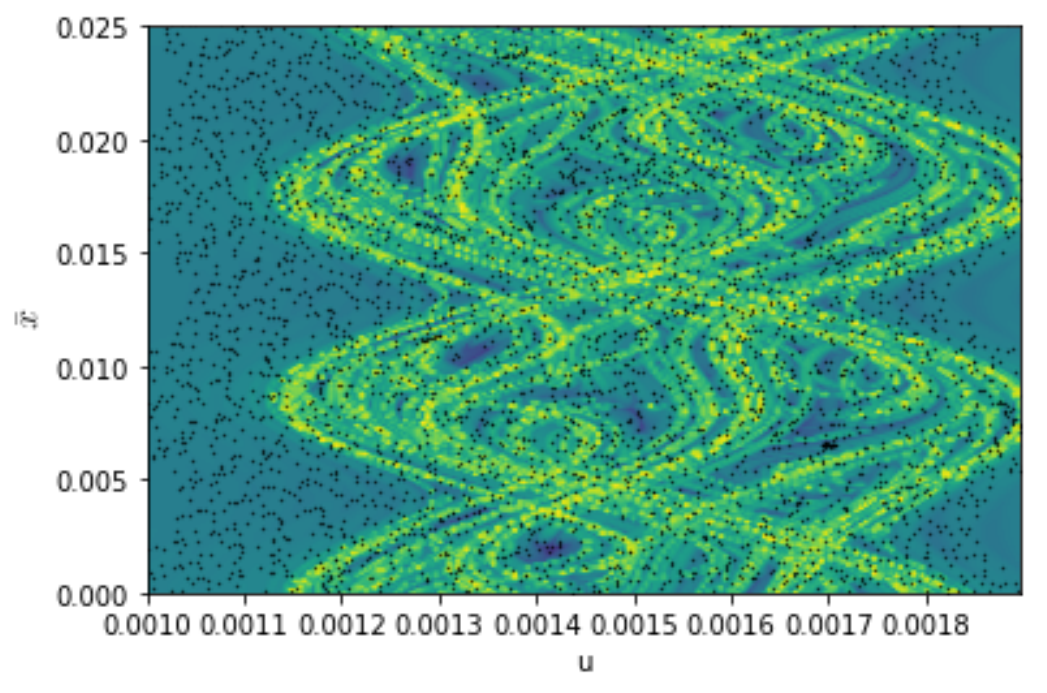}
\caption{(Color online) Contour plots of the FTLE value obtained either with a forward or with a backward time integration ($\Delta \tau = \pm 20$) at $\tau = 700$ \emphpanel{(top left-hand panel)}, $1100$ \emphpanel{(top right-hand)}, $1500$ \emphpanel{(bottom left-hand)}, $2100$ \emphpanel{(bottom right-hand)}. Regions with high FTLE are represented in yellow (color scheme from blue to yellow). Passive tracers are marked in black.
\label{fig_multimodeFTLE}}
\end{figure}
We first characterize phase-space dynamics in the presence of one mode with resonance velocity \(u_r=0.0015\). In \figref{fig_singlemodeFTLE}, we show attractive and repulsive transport barriers (in blue and red, respectively) associated with a time span of $\Delta \tau = 20$ together with a collection of tracers at two different snapshots, \ie $\tau = 700$ and $800$, to underline the saturation dynamics (cf. \figref{fig_nlnlnl}). Yellow lines are placed at $u_{r} \pm \Delta u^{c}_{NL}$ and $u_{r} \pm \Delta u_{NL}$, as indicated in the figure caption. As expected, the phase space is divided into different macro-regions with negligible exchange of particles between them representing the inside and the outside, respectively, of a clump which moves coherently, \ie only with just minor deformation of its structure. This dynamics is confirmed by the evolution of the tracers and can be attributed to the relatively low-amplitude oscillations of the electric field around its saturation value. In particular, as shown in \citet{CFMZJPP}, greater amplitude oscillations of the scalar potential can de-trap a fraction of the particles originally inside the clump and vice versa. The value of $\Delta u^{c}_{NL}$ is consistent with the maximum velocity width of the region enclosed by the LCS. Tracers outside $u_{r} \pm \Delta u_{NL}$ have been colored in order to highlight their dynamics. It can be argued how these particles are exchanging, on average, less power with respect to those ones in the inside region which are moving towards the x-point consistent with the already introduced resonance width.

The same technique can be applied to describe the resonance overlap process introduced in Sec. \ref{overlap}. In \figref{fig_multimodeFTLE} (cf. the evolution of the distribution function in \figref{fig_nlnlnl_}) we show, for each point of the phase space, the largest FTLE value obtained by either a forward or a backward time integration with \(\Delta \tau = \pm 20\). We plot, for the sake of clarity, attractive and repulsive LCSs with the same color since this difference is not relevant for the present analysis. For $\tau=700$, three closed domains are depicted, while trajectories characterized by \(u \lesssim 0.0013\) and \(u \gtrsim 0.0017\) are only slightly deformed. The three resonances do not behave independently, thus generating overlap at saturation time as described in the previous Section and the LCS technique allows one to illuminate this dynamics at each time step. In particular, in the first panel of \figref{fig_multimodeFTLE}, even if the clumps look undistorted and the scalar potentials have approximately single-mode simulation values, the FTLE profile shows peaked structures in the region between adjacent resonances, \ie \(u_{r} \sim 0.0015\) and \(u_{r} \sim 0.0017\), thus suggesting transport processes in the phase space (for a detailed description of this process, see \citet{CFMZJPP}). The overlap is more evident in the next two snapshots ($\tau=1100$ and $1500$), where this region is filled by a convoluted tangle of repulsive and attractive structures. In the last snapshot ($\tau=2100$), the overlap is complete and a unique structure is formed consistent with the formation of a plateau in the particle distribution function.

\section{Mode amplitude at saturation}\label{modes}
The case analyzed in the previous Sections corresponds to an intermediate situation between the limiting cases of isolated and strongly overlapping resonances. The \emphpanel{left-hand panel} of \figref{fig_over} shows that the mode saturation levels are larger in the multi-mode simulations than in the single-mode runs. This feature of the BPS has been discussed in \citet{BB95a} (see also references therein), where sources and sink are included in the model. Due to the enhancement of the distribution function plateau in the overlapping regime, the amount of free energy of the interaction process becomes larger, and this is the cause of the observed mode saturation level.

In the following, we numerically show the overlapping saturation as a function of the phase-velocity separation of neighboring modes, highlighting the feature mentioned above in the collisionless case. Equivalent to the conclusion drawn in \citet{BB95a}, we provide a visual interpretation of this process in terms of the areas relative to the distribution function distortions, \ie the particle number. In particular, we initialize the beam particles using a distribution function with a positive constant gradient and $\eta=0.002$, \ie we consider a linear initial profile in the velocity space. We run 7 simulations with two modes, fixing one of the resonances (namely at $u_{r1}=0.00135$) and sweeping the other one (dubbed $u_{r2}$) in order to span, with constant velocity increment, the resonance separations $\Delta u_{SEP}\equiv u_{r2}-u_{r1}$ from $5\times10^{-6}$ to $2\times10^{-3}$. For each case, we compare numerical results with the evolutions obtained for isolated resonances.

In the presence of multiple modes and resonances, we recall that the total momentum and energy are conserved. In particular, total conserved momentum can be written as \citep{OWM71,CFMZJPP}
\begin{align}\label{momcons}
\mathcal{K}_P=\sum_{\ell}\frac{|\bar{\phi}_{\ell}|^{2}}{\ell}+\frac{2}{\eta N}\sum_i u_i\;.
\end{align}
\begin{figure}
\centering
\includegraphics[width=0.34\textwidth]{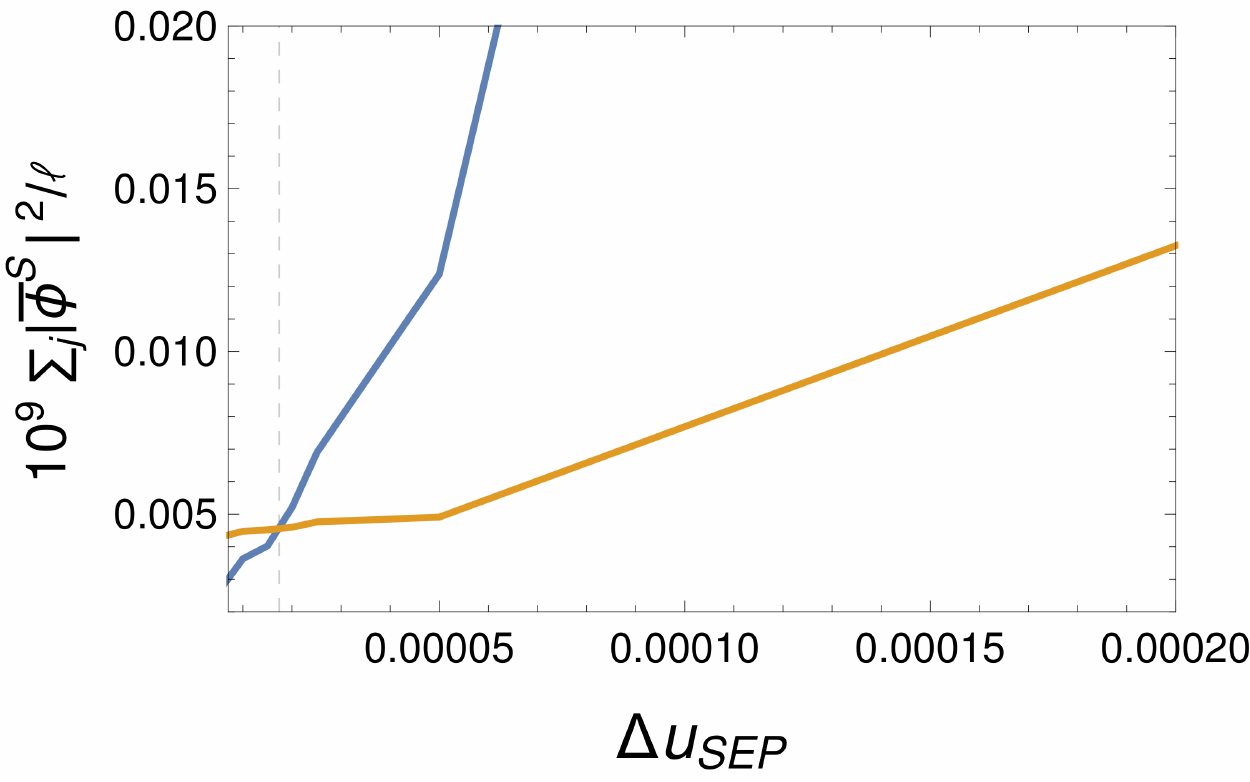}
\includegraphics[width=0.32\textwidth]{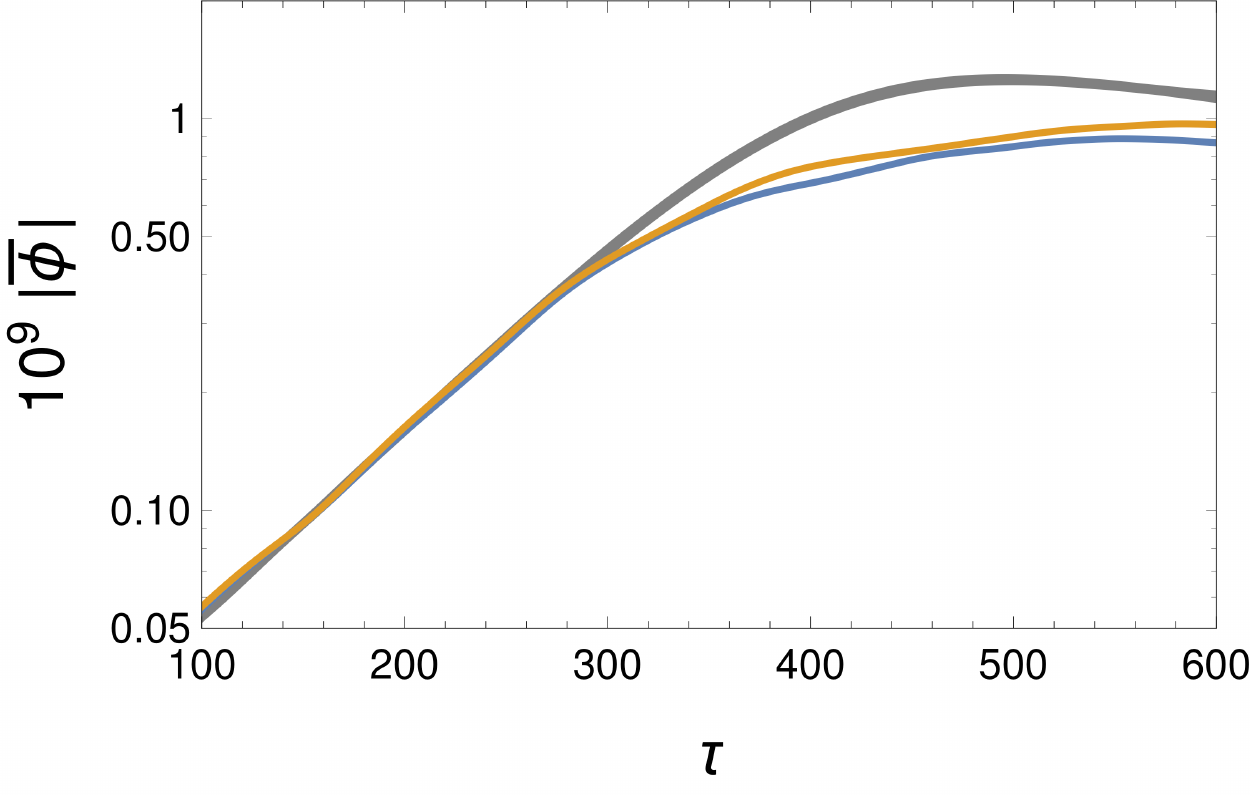}
\includegraphics[width=0.32\textwidth]{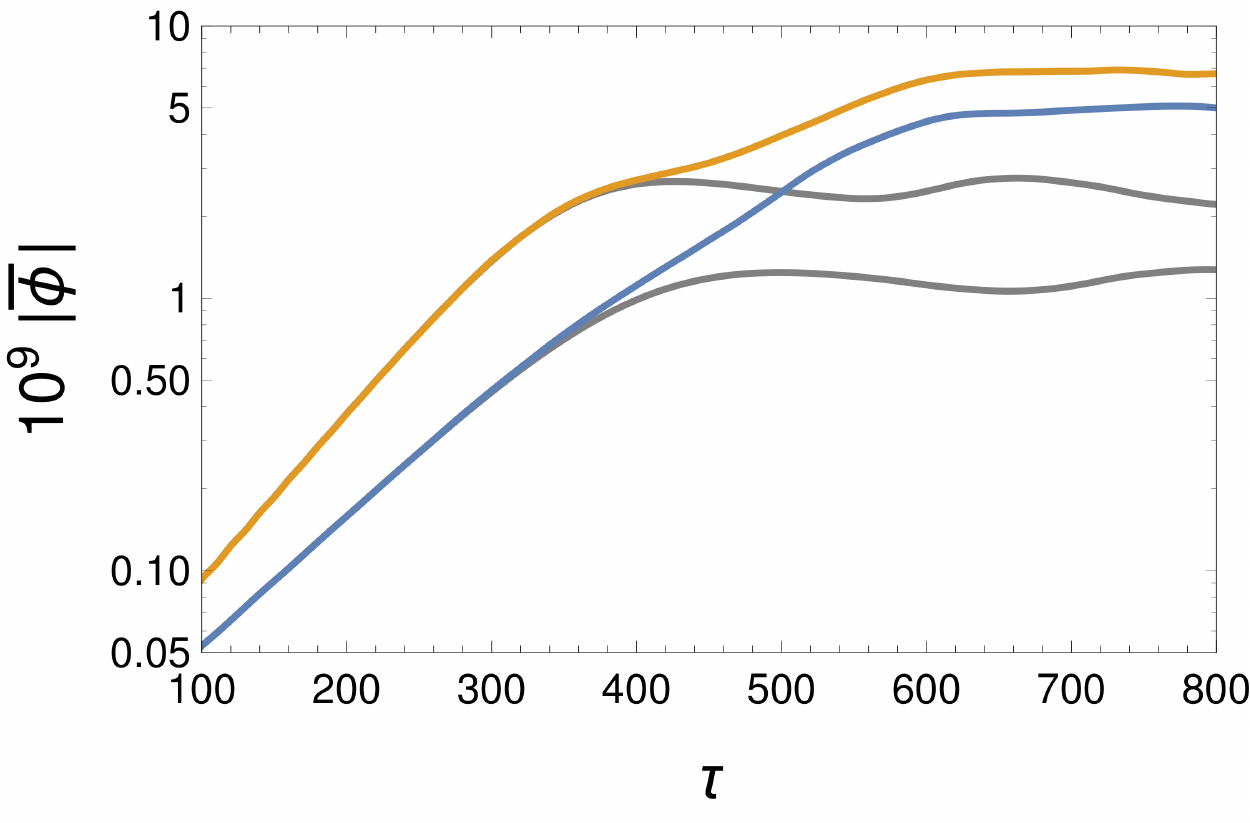}
\caption{(Color online) Left-hand panel: Saturation level $\sum_\ell|\bar{\phi}_{\ell}^S|^{2}/\ell$ as a function of $\Delta u_{SEP}=u_{r2}-u_{r1}$ for self-consistent simulations of two resonances (blue) compared with the saturation level obtained artificially superposing the evolution of the individual isolated resonances (yellow). The threshold for the onset for enhanced saturation level is the intersection point (dashed gray line) $\Delta u^*_{SEP}\simeq 1.75\times10^{-5}$. Center and right-hand panels: Time evolution of the modes for the multi-mode (colored lines) and single-mode (grey lines) systems, for a case below threshold, \ie $\Delta u_{SEP}\simeq 5.5\times10^{-6}$ \emphpanel{(left-hand panel)} and above threshold, \ie $\Delta u_{SEP}\simeq 2\times10^{-4}$ \emphpanel{(right-hand panel)}.
\label{fig_satu}}
\end{figure}
Thus, for each case, we can measure the saturation level as $\sum_\ell|\bar{\phi}_{\ell}^S|^{2}/\ell$. In \figref{fig_satu} \emphpanel{(left-hand panel)}, we plot an interpolation of this quantity as function of $\Delta u_{SEP}$ for the self-consistent simulations of two modes (blue line) compared with the saturation levels obtained artificially superposing the evolution of isolated resonances (yellow line). Note that for vanishing resonance separation, as expected, the self-consistent saturation level is half of the value obtained by the artificial addition of single isolated modes. In fact, for coalescing resonances, the modes become different realizations of the same fluctuating field and their saturation amplitude is reduced by a factor of two. It is immediately seen that a threshold value ($\Delta u^*_{SEP}\simeq 1.75\times10^{-5}$) emerges, below which the saturation level for the self-consistent evolution is lower than that obtained by artificial superposition of single isolated modes: this corresponds to the regime of strongly overlapped resonances. The opposite situation occurs above $\Delta u^*_{SEP}$ (clearly provided that $\Delta u_{SEP}\lesssim 2\Delta u^f_{NL}$, otherwise resonances are not overlapped and the modes evolve as single isolated fluctuations). In order to better illustrate such a behavior, in \figref{fig_satu}, we also plot the single- and multi-mode evolutions, for a case below \emphpanel{(center panel)} and above (\emphpanel{right-hand panel)} threshold, respectively.

\subsection{Interpretative model}
When the multi-mode saturation level is larger than for single isolated modes, it is evident that a more efficient process can tap energy from the particle phase space. Following the line of \citet{BB95a}, we discuss a toy model to qualitatively describe this effect and the threshold condition introduced above. 

\begin{figure}
\centering
\includegraphics[width=0.45\textwidth]{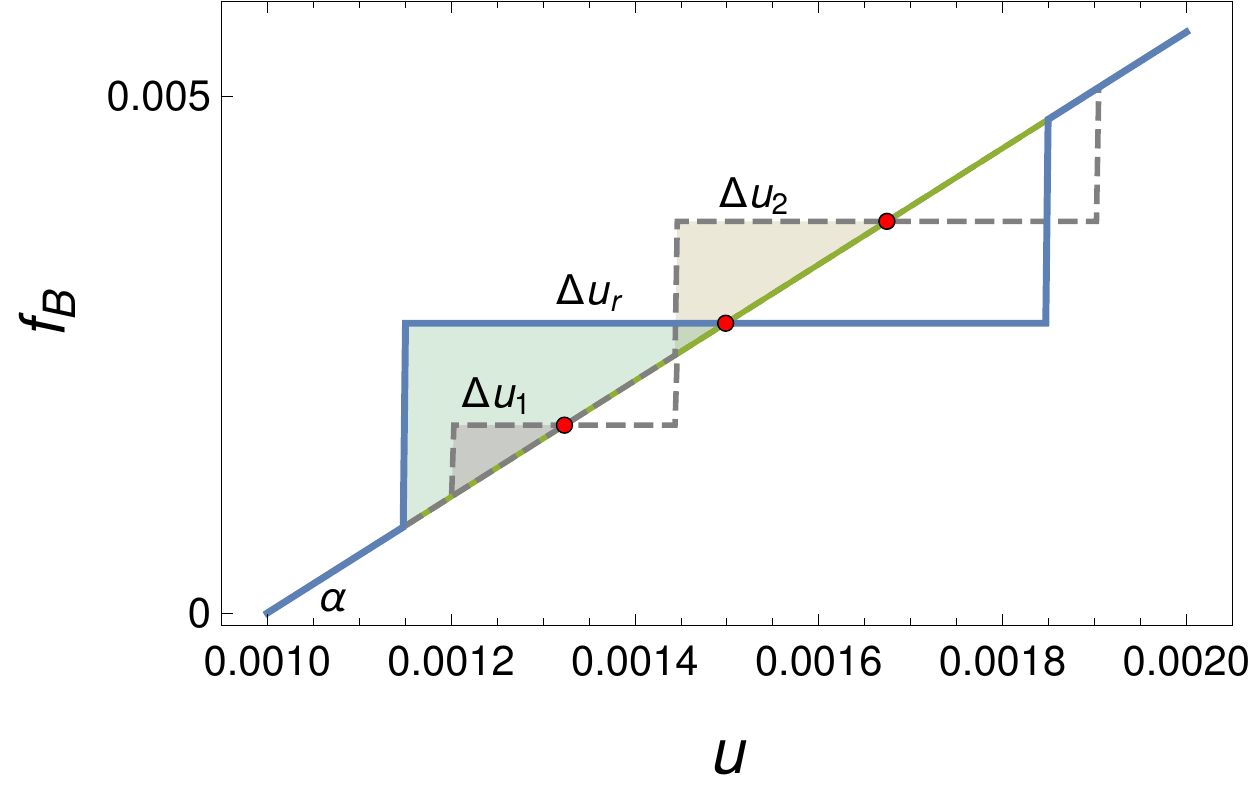}
\includegraphics[width=0.42\textwidth]{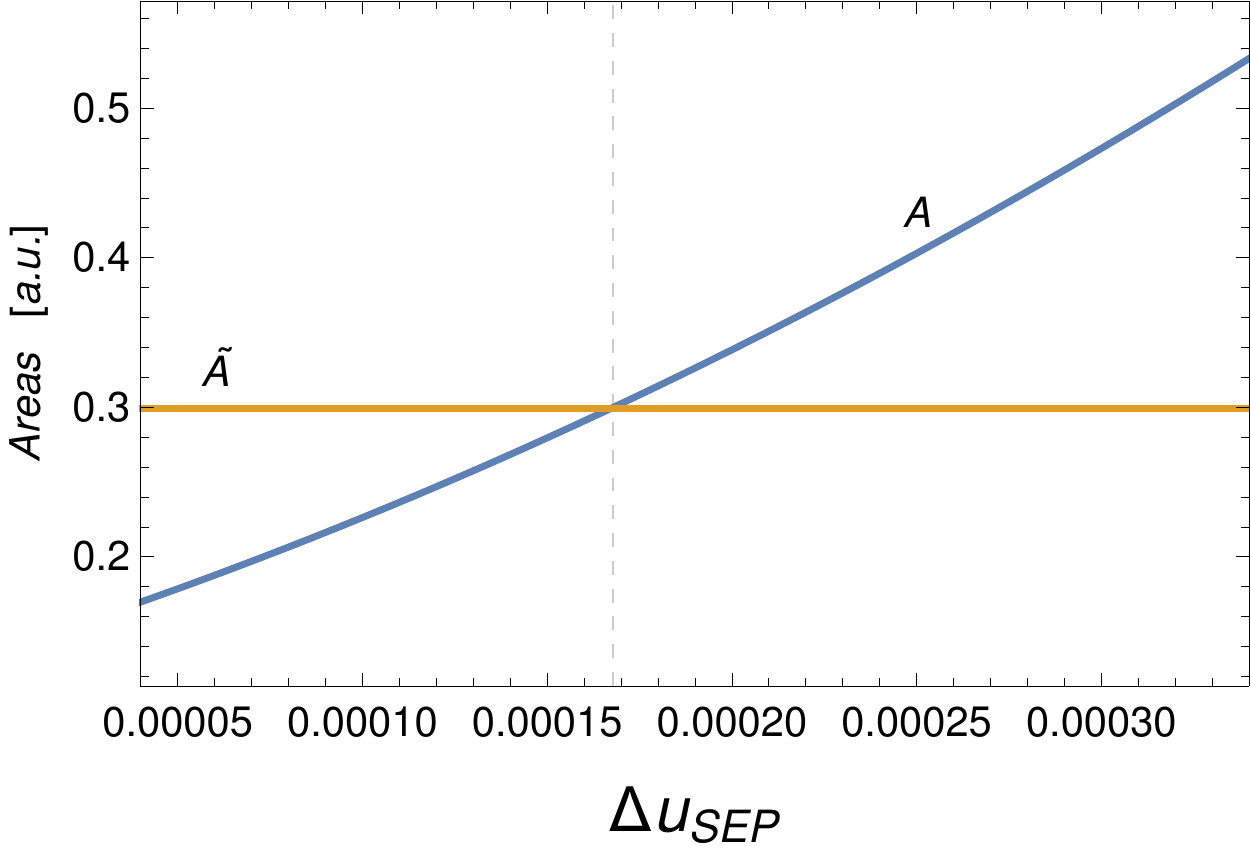}
\caption{(Color online) Left-hand panel: Representation for the distribution function at saturation in the case of single isolated modes (grey dashed lines) and of one overlapped resonance (blue solid line) modeled using \erefs{hfdjksiulh} and \reff{kjshhh}. Red dots correspond, from left to right, to $u_1$, $u_r$ and $u_2$, respectively. Corresponding velocity spreads are indicated in the plot. Right-hand panel: Behavior of $\tilde{A}$ (yellow) and $A$ for $\rho=1$ (blue) in terms of $\Delta u_{SEP}$. Dashed gray line indicates the intersection point.
\label{fig_satumo}}
\end{figure}

For each resonance, respectively called $u_1$ and $u_2$, we assume to model the non-linear distorted distribution function at saturation by a flattening over a certain region; that is, as horizontal lines centered at the resonance position and extended over twice the non-linear velocity spread, respectively denoted as $\Delta u_1$ and $\Delta u_2$ for the two considered resonances. As described in the previous Sections, resonance overlap occurs when the flattening regions intersect. We then describe the overlapping resonances  as a single one having a new resonance velocity and non-linear velocity spread defined, respectively, by
\begin{align}\label{hfdjksiulh}
u_r=u_1+\Delta u_{SEP}/2\;,\\
\Delta u_r=\rho((u_2+\Delta u_2)-(u_1-\Delta u_1))/2\;,\label{kjshhh}
\end{align}
where $\Delta u_{SEP}=u_2-u_1$ and $\rho=\mathcal{O}(1)$ is a control parameter for the model. This scheme is illustrated, for $\rho=1$, in \figref{fig_satumo} \emphpanel{(left-hand panel)}, where we used realistic quantities (estimated from $\Delta u_{NL}$ ) for comparison with the simulation results described above.

The non-linear mode saturation corresponds to particle transfer from large to small velocities as indicated in the momentum conservation law \eref{momcons}. Thus, the relevant quantities to be analyzed are the areas of the triangles that are colored in the figure. These correspond to the amount of available free energy, in the interpretation of \citet{BB95a}. Evolving the two modes as single isolated resonances, the saturation level scales as
\begin{align}
\tilde{A}= 
\tan\alpha((\Delta u_1)^2+(\Delta u_2)^2)/2\;,
\end{align}
\ie as the sum of the areas built around the two resonances. Here $\alpha$ denotes the angular coefficient of the initial linear distribution. Meanwhile, the area of the triangle corresponding to the new merged resonance can be evaluated by means of \erefs{hfdjksiulh} and \reff{kjshhh} as
\begin{align}
A=\rho^2 \tan\alpha(\Delta u_{SEP}+\Delta u_1+\Delta u_2)^2/8\;.
\end{align}
Assuming to neglect the velocity dependence of $\Delta u_2$ (as $u_2$ is swept to change $\Delta u_{SEP}$), \figref{fig_satumo} \emphpanel{(right-hand panel)} shows $\tilde{A}$ and $A$ (for $\rho=1$) as functions of the resonance separation.

For $\Delta u_{SEP} > \Delta u_1 + \Delta u_2$ there is no resonance overlap and the system response is consistent with two isolated resonances. For decreasing $\Delta u_{SEP}$ below the onset of the overlap (\ie the system depicted in \figref{fig_satumo}, \emphpanel{left-hand panel}), the saturation level for the merged resonance is larger than for the single isolated modes, \ie $A>\tilde{A}$, or, more precisely, $A=2\tilde{A}$ at the onset condition, suggesting a sudden transition in the non-linear dynamics due to the synergistic behavior of the interacting modes. In particular, it is readily verified that $A>\tilde{A}$ as long as a lower critical threshold is exceeded (as in the simulation results), namely
\begin{align}
\Delta u_{SEP}^*=-\Delta u_1-\Delta u_2+2\sqrt{((\Delta u_1)^2+(\Delta u_2)^2)/\rho^2}\;.
\end{align}
Thus, we can identify a range for enhanced or synergistic interaction of overlapping resonances
\begin{align}
\Delta u^*_{SEP}<\Delta u_{SEP}<\Delta u_1 + \Delta u_2\;.
\end{align}
Meanwhile, for $\Delta u_{SEP}<\Delta u^*_{SEP}$, the two resonances are strongly overlapping and $A<\tilde{A}$, with $A=\tilde{A}/2$ in the limit of $\Delta u_{SEP}=0$. This is consistent with our numerical simulations.

The discrepancy between the intersection point ($\simeq 1.7\times10^{-4}$) of \figref{fig_satumo} \emphpanel{(right-hand panel)} with respect to the value found in numerical simulations shown in \figref{fig_satu} \emphpanel{(left-hand panel)} is due to intrinsic details of the non-linear evolution and the simplification contained in the model: it can be taken into account by properly setting the model control parameter $\rho$. Indeed, a proper match of the model theoretical threshold with numerical simulations can be obtained using $\rho=1.41$. As a result, the present model shows how the mode saturation level can be described via the analysis of wave-particle power transfer in phase space. The model can also be made quantitative (and, thus, predictive) by a proper choice of the control parameter $\rho$.

\section{Concluding remarks}\label{conclusion}
The BPS and its paradigmatic implications in plasma and fusion physics have been widely discussed across the literature. 
The present analysis focused on the refinement of some subtle questions mandatory for practical applications, in order to make quantitative estimates of the non-linear wave-particle interaction.

A study of the non-linear velocity spread allowed selection of a proper region in the velocity space characterizing the resonance width. Analyzing the simulations of isolated resonances up to the limiting distance of their overlap, we show that also the ``tails'' around the plateau play an important dynamical role. It is worth noting that such regions do not contain trapped particles. Nonetheless, their existence must be carefully taken into account when assessing the separation of two resonant modes. In fact, the plateau region, mainly coinciding with the clump size in the phase space, would underestimate the transport of particles between two adjacent resonances due to the important role played by un-trapped particles. The shape of the clumps and phase-space dynamics have been described using the LCS technique, which clarifies how particles outside the effective resonance width exchange limited power with the mode spectrum. 

Finally, we study the enhanced saturation of interacting resonant modes with respect to the level of isolated resonances. We quantitatively describe the merging/overlap of two adjacent resonances. A critical distance exists in the velocity space, above which the two resonances are isolated and below which they are too overlapped to be really distinct in the velocity space. In either case, the two fluctuations behave as individual modes and the wave particle power exchange is limited. Only when the two resonances are adjacent and the power transfer from particles to modes is maximized by an enforced velocity spread (essentially the sum of the original ones), can we predict the enhanced saturation observed for instance in \citet{spb16} and \citet{vnf18}.

The analysis of this paper offers a summary of the relevant features of the bump-on-tail paradigm (to which the BPS is isomorphic), in view of its paradigmatic character in describing many several important physical problems. Quantitative prediction of resonance overlap and of enhanced saturation conditions and levels was the original motivation of this work, which has led us to conclude that the system evolution is strongly influenced by the global features of the distribution function.

\vspace{1cm}

{\small ** We would like to gratefully thank Fulvio Zonca for having inspired this work and for his valuable suggestions. NC would like to thank Philip Lauber and Thomas Hayward-Schneider for their fruitful discussions and advices. 

This work has been carried out within the framework of the EUROfusion Consortium [Enabling Research Projects: NAT (AWP17-ENR-MFE-MPG-01), MET (AWP19-ENR-01-ENEA-05)] and has received funding from the Euratom research and training programme 2014-2018 and 2019-2020 under grant agreement No 633053. The views and opinions expressed herein do not necessarily reflect those of the European Commission. **}


\begin{thebibliography}{53}
\expandafter\ifx\csname natexlab\endcsname\relax\def\natexlab#1{#1}\fi
\def\au#1{#1} \def\ed#1{#1} \def\yr#1{#1}\def\at#1{#1}\def\jt#1{\textit{#1}}
  \def\bt#1{#1}\def\bvol#1{\textbf{#1}} \def\vol#1{#1} \def\pg#1{#1}
  \def\publ#1{#1}\def\arxiv#1{#1}\def\org#1{#1}\def\st#1{\textit{#1}}

\bibitem[{Al'Tshul'} \& {Karpman}(1966)]{AK66}
{\sc \au{{Al'Tshul'}, L.~M.} \& \au{{Karpman}, V.~I.}} \yr{1966}  \at{Theory of
  nonlinear oscillations in a collisionless plasma}.  \jt{Sov. Phys. JEPT}
  \bvol{22}~(2),  \pg{361--369}.

\bibitem[{Berk} \& {Breizman}(1994)]{BB94b}
{\sc \au{{Berk}, H.L.} \& \au{{Breizman}, B.N.}} \yr{1994}  \bt{{Scenarios for
  the nonlinear evolution of beam-driven instability with a weak source}}. \jt{AIP Conf. Series},  \bvol{314},  \pg{pp. 140--155}.

\bibitem[{Berk} {\em et~al.\/}(1994){Berk}, {Breizman} \& {Pekker}]{BB94a}
{\sc \au{{Berk}, H.L.}, \au{{Breizman}, B.N.} \& \au{{Pekker}, M.}} \yr{1994}
  \at{{Basic principles approach for studying nonlinear Alfv{\'e}n wave-alpha
  particle dynamics}}.  \jt{AIP Conf. Series}  \bvol{311},  \pg{18--31}.

\bibitem[{Berk} {\em et~al.\/}(1992){Berk}, {Breizman} \& {Ye}]{BB92PRL}
{\sc \au{{Berk}, H.L.}, \au{{Breizman}, B.N.} \& \au{{Ye}, H.}} \yr{1992}
  \at{{Scenarios for the nonlinear evolution of alpha-particle-induced
  Alfv{\'e}n wave instability}}.  \jt{Phys. Rev. Lett.}  \bvol{68}~(24),
  \pg{3563--3566}.

\bibitem[{Berk} \& {Breizman}(1990{\natexlab{{\em a\/}}})]{BB90a}
{\sc \au{{Berk}, H.~L.} \& \au{{Breizman}, B.~N.}} \yr{1990{\natexlab{{\em
  a\/}}}}  \at{{Saturation} of a single mode driven by an energetic injected
  beam. {I}. {Plasma} wave problem}.  \jt{Phys. Fluids B}  \bvol{2}~(9),
  \pg{2226--2234}.

\bibitem[{Berk} \& {Breizman}(1990{\natexlab{{\em b\/}}})]{BB90b}
{\sc \au{{Berk}, H.~L.} \& \au{{Breizman}, B.~N.}} \yr{1990{\natexlab{{\em
  b\/}}}}  \at{{Saturation} of a single mode driven by an energetic injected
  beam. {II.} {Electrostatic} ``universal'' destabilization mechanism}.
  \jt{Phys. Fluids B}  \bvol{2}~(9),  \pg{2235--2245}.

\bibitem[{Berk} \& {Breizman}(1990{\natexlab{{\em c\/}}})]{BB90c}
{\sc \au{{Berk}, H.~L.} \& \au{{Breizman}, B.~N.}} \yr{1990{\natexlab{{\em
  c\/}}}}  \at{{Saturation} of a single mode driven by an energetic injected
  beam. {III.} {Alfv{\'e}n} wave problem}.  \jt{Phys. Fluids B}  \bvol{2}~(9),
  \pg{2246--2252}.

\bibitem[{Berk} {\em et~al.\/}(1995{\natexlab{{\em a\/}}}){Berk}, {Breizman},
  {Fitzpatrick} \& {Wong}]{BB95b}
{\sc \au{{Berk}, H.~L.}, \au{{Breizman}, B.~N.}, \au{{Fitzpatrick}, J.} \&
  \au{{Wong}, H.~V.}} \yr{1995{\natexlab{{\em a\/}}}}  \at{{Line} broadened
  quasi-linear burst model [fusion plasma]}.  \jt{Nucl. Fusion}
  \bvol{35}~(12),  \pg{1661--1668}.

\bibitem[{Berk} {\em et~al.\/}(1995{\natexlab{{\em b\/}}}){Berk}, {Breizman} \&
  {Pekker}]{BB95a}
{\sc \au{{Berk}, H.~L.}, \au{{Breizman}, B.~N.} \& \au{{Pekker}, M.}}
  \yr{1995{\natexlab{{\em b\/}}}}  \at{Numerical simulation of bump-on-tail
  instability with source and sink}.  \jt{Phys. Plasmas}  \bvol{2}~(8),
  \pg{3007--1259}.

\bibitem[{Breizman} {\em et~al.\/}(1993){Breizman}, {Berk} \& {Ye}]{BB93}
{\sc \au{{Breizman}, B.~N.}, \au{{Berk}, H.~L.} \& \au{{Ye}, H.}} \yr{1993}
  \at{Collective transport of alpha particles due to alfv{\'e}n wave
  instability}.  \jt{Phys. Fluids B}  \bvol{5}~(9),  \pg{3217--3226}.

\bibitem[{Breizman} \& {Sharapov}(2011)]{BS11}
{\sc \au{{Breizman}, B.~N.} \& \au{{Sharapov}, S.~E.}} \yr{2011}  \at{{Major}
  minority: energetic particles in fusion plasmas}.  \jt{Plasma Phys. Contr.
  Fusion}  \bvol{53}~(5),  \pg{054001}.

\bibitem[{Carlevaro} {\em et~al.\/}(2015){Carlevaro}, {Falessi}, {Montani} \&
  {Zonca}]{CFMZJPP}
{\sc \au{{Carlevaro}, N.}, \au{{Falessi}, M.~V.}, \au{{Montani}, G.} \&
  \au{{Zonca}, F.}} \yr{2015}  \at{Nonlinear physics and energetic particle
  transport features of the beam-plasma instability}.  \jt{J. Plasma Phys.}
  \bvol{81}~(5),  \pg{495810515}.

\bibitem[{Carlevaro} {\em et~al.\/}(2014){Carlevaro}, {Fanelli}, {Garbet},
  {Ghendrih}, {Montani} \& {Pettini}]{CFGGMP14}
{\sc \au{{Carlevaro}, N.}, \au{{Fanelli}, D.}, \au{{Garbet}, X.},
  \au{{Ghendrih}, P.}, \au{{Montani}, G.} \& \au{{Pettini}, M.}} \yr{2014}
  \at{Beam-plasma instability and fast particles: the {Lynden}-{Bell}
  approach}.  \jt{Plasma Phys. Contr. Fusion}  \bvol{56}~(3),  \pg{035013}.

\bibitem[{Carlevaro} {\em et~al.\/}(2019{\natexlab{{\em a\/}}}){Carlevaro},
  {Finelli} \& {Montani}]{finelli19}
{\sc \au{{Carlevaro}, N.}, \au{{Finelli}, F.} \& \au{{Montani}, G.}}
  \yr{2019{\natexlab{{\em a\/}}}}  \at{Reanalysis of the beam-plasma
  instability using the {Dyson-like} equation formalism}.  \jt{Europhys. Lett.}
   \bvol{127},  \pg{25002}.

\bibitem[{Carlevaro} {\em et~al.\/}(2016{\natexlab{{\em a\/}}}){Carlevaro},
  {Milovanov}, {Falessi}, {Montani}, {Terzani} \& {Zonca}]{ncentropy}
{\sc \au{{Carlevaro}, N.}, \au{{Milovanov}, A.~V.}, \au{{Falessi}, M.~V.},
  \au{{Montani}, G.}, \au{{Terzani}, D.} \& \au{{Zonca}, F.}}
  \yr{2016{\natexlab{{\em a\/}}}}  \at{Mixed diffusive-convective relaxation of
  a warm beam of energetic particles in cold plasma}.  \jt{Entropy}
  \bvol{18}~(4),  \pg{143}.

\bibitem[{Carlevaro} {\em et~al.\/}(2016{\natexlab{{\em b\/}}}){Carlevaro},
  {Montani}, {Wang} \& {Zonca}]{nceps16}
{\sc \au{{Carlevaro}, N.}, \au{{Montani}, G.}, \au{{Wang}, X.} \& \au{{Zonca},
  F.}} \yr{2016{\natexlab{{\em b\/}}}}  \at{Hamiltonian bump-on-tail model:
  interpretation of {EP}/{AE} interaction}.  \jt{\emph{In} 43rd EPS Conference
  on Plasma Physics}  \bvol{40A},  \pg{P5.018}.

\bibitem[{Carlevaro} {\em et~al.\/}(2018){Carlevaro}, {Montani} \&
  {Zonca}]{nceps18}
{\sc \au{{Carlevaro}, N.}, \au{{Montani}, G.} \& \au{{Zonca}, F.}} \yr{2018}
  \at{Resonance overlap and non-linear velocity spread in {Hamiltonian}
  beam-plasma systems}.  \jt{\emph{In} 45th EPS Conference on Plasma Physics}
  \bvol{42A},  \pg{P5.1067}.

\bibitem[{Carlevaro} {\em et~al.\/}(2019{\natexlab{{\em b\/}}}){Carlevaro},
  {Montani}, {Zonca}, {Lauber} \& {Hayward-Schneider}]{nceps19}
{\sc \au{{Carlevaro}, N.}, \au{{Montani}, G.}, \au{{Zonca}, F.}, \au{{Lauber},
  P.} \& \au{{Hayward-Schneider}, T.}} \yr{2019{\natexlab{{\em b\/}}}}
  \at{Beam-plasma system as reduced model for {ITER} relevant energetic
  particle transport}.  \jt{\emph{In} 46th EPS Conference on Plasma Physics}
  \bvol{43C},  \pg{P5.1014}.

\bibitem[{Chen} \& {Zonca}(2016)]{ZCrmp}
{\sc \au{{Chen}, L.} \& \au{{Zonca}, F.}} \yr{2016}  \at{Physics of {Alfv\'en}
  waves and energetic particles in burning plasmas}.  \jt{Rev. Mod. Phys.}
  \bvol{88}~(1),  \pg{015008}.

\bibitem[{Chirikov}(1960)]{Ch60}
{\sc \au{{Chirikov}, B.V.}} \yr{1960}  \at{Resonance processes in magnetic
  traps}.  \jt{J. Nucl. Energy - Part C Plasma Phys.}  \bvol{1},  \pg{263--260}.

\bibitem[{Chirikov}(1979)]{Ch79}
{\sc \au{{Chirikov}, B.~V.}} \yr{1979}  \at{A universal instability of
  many-dimensional oscillator systems}.  \jt{Phys. Rept.}  \bvol{52}~(5),
  \pg{263--379}.

\bibitem[Di~Giannatale {\em et~al.\/}(2018{\natexlab{{\em a\/}}})Di~Giannatale,
  Falessi, Grasso, Pegoraro \& Schep]{Di_Giannatale_2018}
{\sc \au{Di~Giannatale, G.}, \au{Falessi, M.~V.}, \au{Grasso, D.},
  \au{Pegoraro, F.} \& \au{Schep, T.~J.}} \yr{2018{\natexlab{{\em a\/}}}}
  \at{Coherent transport structures in magnetized plasmas. i. {Theory}}.
  \jt{Phys. Plasmas}  \bvol{25}~(5),  \pg{052306}.

\bibitem[Di~Giannatale {\em et~al.\/}(2018{\natexlab{{\em b\/}}})Di~Giannatale,
  Falessi, Grasso, Pegoraro \& Schep]{Di_Giannatale_2018b}
{\sc \au{Di~Giannatale, G.}, \au{Falessi, M.~V.}, \au{Grasso, D.},
  \au{Pegoraro, F.} \& \au{Schep, T.~J.}} \yr{2018{\natexlab{{\em b\/}}}}
  \at{Coherent transport structures in magnetized plasmas. ii. {Numerical
  results}}.  \jt{Phys. Plasmas}  \bvol{25}~(5),  \pg{052307}.

\bibitem[{Drummond} \& {Pines}(1962)]{Pines}
{\sc \au{{Drummond}, W.~E.} \& \au{{Pines}, D.}} \yr{1962}  \at{Non-linear
  stability of plasma oscillations}.  \jt{Nucl. Fusion Suppl. Part.}  \bvol{3},
   \pg{1049--1057}.

\bibitem[{Elskens} \& {Escande}(2003)]{EEbook}
{\sc \au{{Elskens}, Y.} \& \au{{Escande}, D.~F.}} \yr{2003} {\em Microscopic
  Dynamics of Plasmas Chaos\/}.  \publ{Taylor Francis Ltd}.

\bibitem[{Esarey} {\em et~al.\/}(1996){Esarey}, {Sprangle}, {Krall} \&
  {Ting}]{ES96}
{\sc \au{{Esarey}, E.}, \au{{Sprangle}, P.}, \au{{Krall}, J.} \& \au{{Ting},
  A.}} \yr{1996}  \at{Overview of plasma-based accelerator concepts}.  \jt{IEEE
  Tran. Plasma Science}  \bvol{24}~(2),  \pg{252--288}.

\bibitem[{Escande} {\em et~al.\/}(2018){Escande}, {B\`enisti}, {Elskens},
  {Zarzoso} \& {Doveil}]{ee18}
{\sc \au{{Escande}, D.F.}, \au{{B\`enisti}, D.}, \au{{Elskens}, Y.},
  \au{{Zarzoso}, D.} \& \au{{Doveil}, D.}} \yr{2018}  \at{Basic microscopic
  plasma physics from {N-body} mechanics}.  \jt{Rev. Mod. Plasma Phys.}
  \bvol{2},  \pg{9}.

\bibitem[{Escande} \& {Doveil}(1981)]{ED81}
{\sc \au{{Escande}, D.F.} \& \au{{Doveil}, F.}} \yr{1981}  \at{Renormalization
  method for computing the threshold of the large-scale stochastic instability
  in two degrees of freedom {Hamiltonian} systems}.  \jt{J. Stat. Phys.}
  \bvol{26}~(2),  \pg{257--284}.

\bibitem[{Falessi} {\em et~al.\/}(2015){Falessi}, {Pegoraro} \& {Schep}]{MPJPP}
{\sc \au{{Falessi}, M.~V.}, \au{{Pegoraro}, F.} \& \au{{Schep}, T.~J.}}
  \yr{2015}  \at{Lagrangian coherent structures and plasma transport
  processes}.  \jt{J. Plasma Phys.}  \bvol{81}~(5),  \pg{495810505}.

\bibitem[{Fried} {\em et~al.\/}(1971){Fried}, {Liu}, {Means} \&
  {Sagdeev}]{ucla_fried}
{\sc \au{{Fried}, B.D.}, \au{{Liu}, C.S.}, \au{{Means}, R.W.} \& \au{{Sagdeev},
  R.Z.}} \yr{1971}  \at{Nonlinear evolution and saturation of an ustable
  electrostatic wave}.  \jt{UCLA Report}  \bvol{PPG-93}.

\bibitem[{Greene}(1968)]{Gr68}
{\sc \au{{Greene}, J.M.}} \yr{1968}  \at{Two-dimensional measure-preserving
  mappings}.  \jt{J. Math. Phys.}  \bvol{9}~(5),  \pg{760--768}.

\bibitem[Haller(2015)]{Haller_2015}
{\sc \au{Haller, G.}} \yr{2015}  \at{{Lagrangian} {Coherent} {Structures}}.
  \jt{Annual Review of Fluid Mechanics}  \bvol{47}~(1),  \pg{137–162}.

\bibitem[{Jaeger} \& {Lichtenberg}(1972)]{JL72}
{\sc \au{{Jaeger}, E.F.} \& \au{{Lichtenberg}, A.J.}} \yr{1972}  \at{Resonant
  modification and destruction of adiabatic invariants}.  \jt{Ann. Phys.}
  \bvol{71}~(2),  \pg{319--356}.

\bibitem[{Keinigs} \& {Jones}(1987)]{KJ86}
{\sc \au{{Keinigs}, R.} \& \au{{Jones}, M.~E.}} \yr{1987}  \at{Two-dimensional
  dynamics of the plasma wakefield accelerator}.  \jt{Phys. Fluids}
  \bvol{30}~(1),  \pg{252--263}.

\bibitem[{Levin} {\em et~al.\/}(1972){Levin}, {Lyubarski{\v i}}, {Onishchenko},
  {Shapiro} \& {Shevchenko}]{L72}
{\sc \au{{Levin}, M.~B.}, \au{{Lyubarski{\v i}}, M.~G.}, \au{{Onishchenko},
  I.~N.}, \au{{Shapiro}, V.~D.} \& \au{{Shevchenko}, V.~I.}} \yr{1972}
  \at{Contribution to the nonlinear theory of kinetic instability of an
  electron beam in plasma}.  \jt{Sov. Phys. JEPT}  \bvol{35}~(5),
  \pg{898--901}.

\bibitem[{Lichtenberg} \& {Lieberman}(2010)]{LL10}
{\sc \au{{Lichtenberg}, A.J} \& \au{{Lieberman}, M.A.}} \yr{2010} {\em Regular
  and Chaotic Dynamics - {Second} Edition\/}.  \publ{Springer-Verlag}.

\bibitem[{Lifshitz} \& {Pitaevskii}(1976)]{LP81}
{\sc \au{{Lifshitz}, E.~M.} \& \au{{Pitaevskii}, L.~P.}} \yr{1976} {\em Course
  of Theoretical Physics, Volume 10: Physical Kinetics\/}.
  \publ{Butterworth-Heinemann}.

\bibitem[{Litos et al.}(2014)]{Li14}
{\sc \au{{Litos et al.}, M.}} \yr{2014}  \at{High-efficiency acceleration of an
  electron beam in a plasma wakefield accelerator}.  \jt{Nature}
  \bvol{515}~(7525),  \pg{92--95}.

\bibitem[{Mynick} \& {Kaufman}(1978)]{MK78}
{\sc \au{{Mynick}, H.~E.} \& \au{{Kaufman}, A.~N.}} \yr{1978}  \at{Soluble
  theory of nonlinear beam-plasma interaction}.  \jt{Phys. Fluids}  \bvol{21},
  \pg{653--663}.

\bibitem[{O'Neil} \& {Malmberg}(1968)]{OM68}
{\sc \au{{O'Neil}, T.~M.} \& \au{{Malmberg}, J.~H.}} \yr{1968}  \at{Transition
  of the dispersion roots from beam-type to {Landau}-type solutions}.
  \jt{Phys. Fluids}  \bvol{11}~(8),  \pg{1754--1760}.

\bibitem[{O'Neil} {\em et~al.\/}(1971){O'Neil}, {Winfrey} \& {Malmberg}]{OWM71}
{\sc \au{{O'Neil}, T.~M.}, \au{{Winfrey}, J.~H.} \& \au{{Malmberg}, J.~H.}}
  \yr{1971}  \at{Nonlinear interaction of a small cold beam and a plasma}.
  \jt{Phys. Fluids}  \bvol{14}~(6),  \pg{1204--1212}.

\bibitem[Pegoraro {\em et~al.\/}(2019)Pegoraro, Bonfiglio, Cappello,
  Di~Giannatale, Falessi, Grasso \& Veranda]{Pegoraro_2019}
{\sc \au{Pegoraro, F.}, \au{Bonfiglio, D.}, \au{Cappello, S.},
  \au{Di~Giannatale, G.}, \au{Falessi, M.~V.}, \au{Grasso, D.} \& \au{Veranda,
  M.}} \yr{2019}  \at{{Coherent} magnetic structures in self-organized
  plasmas}.  \jt{Plasma Phys. Control. Fusion}  \bvol{61}~(4),  \pg{044003}.

\bibitem[{Pommois} {\em et~al.\/}(2017){Pommois}, {Valentini}, {Pezzi} \&
  {Veltri}]{pommo17}
{\sc \au{{Pommois}, K.}, \au{{Valentini}, F.}, \au{{Pezzi}, O.} \&
  \au{{Veltri}, P.}} \yr{2017}  \at{Slow electrostatic fluctuations generated
  by beam-plasma interaction}.  \jt{Phys. Plasmas}  \bvol{24}~(1),
  \pg{012105}.

\bibitem[{Schneller} {\em et~al.\/}(2016){Schneller}, {Lauber} \&
  {Briguglio}]{spb16}
{\sc \au{{Schneller}, M.}, \au{{Lauber}, Ph.} \& \au{{Briguglio}, S.}}
  \yr{2016}  \at{Nonlinear energetic particle transport in the presence of
  multiple {Alfv\'enic} waves in {ITER}}.  \jt{Plasma Phys. Control. Fusion}
  \bvol{58},  \pg{014019}.

\bibitem[{Shalaby} {\em et~al.\/}(2017){Shalaby}, {Broderick}, {Chang},
  {Pfrommer}, {Lamberts} \& {Puchwein}]{shalaby17}
{\sc \au{{Shalaby}, M.}, \au{{Broderick}, A.~E.}, \au{{Chang}, P.},
  \au{{Pfrommer}, C.}, \au{{Lamberts}, A.} \& \au{{Puchwein}, E.}} \yr{2017}
  \at{Importance of resolving the spectral support of beam-plasma instabilities
  in simulations}.  \jt{ApJ}  \bvol{848},  \pg{81}.

\bibitem[{Tao} {\em et~al.\/}(2017){Tao}, {Zonca} \& {Chen}]{tzc17}
{\sc \au{{Tao}, X.}, \au{{Zonca}, F.} \& \au{{Chen}, L.}} \yr{2017}
  \at{Identify the nonlinear wave-particle interaction regime in rising tone
  chorus generation}.  \jt{Geophys. Res. Lett.}  \bvol{44}~(8),
  \pg{3441--3446}.

\bibitem[{Tennyson} {\em et~al.\/}(1994){Tennyson}, {Meiss} \&
  {Morrison}]{TMM94}
{\sc \au{{Tennyson}, J.~L.}, \au{{Meiss}, J.~D.} \& \au{{Morrison}, P.~J.}}
  \yr{1994}  \at{Self-consistent chaos in the beam-plasma instability}.
  \jt{Physica D}  \bvol{71}~(1-2),  \pg{1--17}.

\bibitem[{Tobita} \& {Omura}(2018)]{tobita18}
{\sc \au{{Tobita}, M.} \& \au{{Omura}, Y.}} \yr{2018}  \at{Nonlinear dynamics
  of resonant electrons interacting with coherent langmuir waves}.  \jt{Phys.
  Plasmas}  \bvol{25}~(3),  \pg{032105}.

\bibitem[{Vedenov} {\em et~al.\/}(1961){Vedenov}, {Velikhov} \&
  {Sagdeev}]{Vedenov}
{\sc \au{{Vedenov}, A.~A.}, \au{{Velikhov}, E.~P.} \& \au{{Sagdeev}, R.~Z.}}
  \yr{1961}  \at{Nonlinear oscillations of rarified plasma}.  \jt{Nucl. Fusion}
   \bvol{1},  \pg{82--100}.

\bibitem[{Vlad} {\em et~al.\/}(2018){Vlad}, {Briguglio}, {Fogaccia}, {Fusco},
  {Di Troia}, {Giovannozzi}, {Wang} \& {Zonca}]{vnf18}
{\sc \au{{Vlad}, G.}, \au{{Briguglio}, S.}, \au{{Fogaccia}, G.}, \au{{Fusco},
  V.}, \au{{Di Troia}, C.}, \au{{Giovannozzi}, E.}, \au{{Wang}, X.} \&
  \au{{Zonca}, F.}} \yr{2018}  \at{Single-n versus multiple-n simulations of
  {Alfv\'enic} modes}.  \jt{Nucl. Fusion}  \bvol{58}~(8),  \pg{082020}.

\bibitem[{Volokitin} \& {Krafft}(2012)]{VK12}
{\sc \au{{Volokitin}, A.} \& \au{{Krafft}, C.}} \yr{2012}  \at{Velocity
  diffusion in plasma waves excited by electron beams}.  \jt{Plasma Phys.
  Control. Fusion}  \bvol{54}~(8),  \pg{085002}.

\bibitem[{Wu} {\em et~al.\/}(1994){Wu}, {Cheng} \& {White}]{wu94}
{\sc \au{{Wu}, Y.}, \au{{Cheng}, C.~Z.} \& \au{{White}, R.~B.}} \yr{1994}
  \at{Alpha particle effects on the internal kink and fishbone modes}.
  \jt{Phys. Plasmas}  \bvol{1},  \pg{3369--3377}.

\bibitem[{Wu} {\em et~al.\/}(1995){Wu}, {White}, {Chen} \& {Rosenbluth}]{wu95}
{\sc \au{{Wu}, Y.}, \au{{White}, R.~B.}, \au{{Chen}, Y.} \& \au{{Rosenbluth},
  M.~N.}} \yr{1995}  \at{Nonlinear evolution of the alpha-particle-driven
  toroidicity-induced {Alfv\'en} eigenmode}.  \jt{Phys. Plasmas}  \bvol{2},
  \pg{4555--4562}.

\end{thebibliography}

\end{document}